\documentclass[aps,prb,twocolumn,amsmath,10pt,nobibnotes,longbibliography,floatfix,a4paper]{revtex4-1}
\usepackage{bm}
\usepackage{enumerate}
\usepackage[hmargin=1.5cm,vmargin=2.7cm]{geometry}
\usepackage{graphicx}
\usepackage{mathptmx}
\usepackage{multirow,bigdelim}
\usepackage{pifont}
\usepackage{textcomp}
\usepackage{titlesec}
\usepackage{hyperref}

\hypersetup{
pdftitle={Deterministic creation of entangled atom-light Schr{\"o}dinger-cat states},
pdfauthor={Bastian Hacker, Stephan Welte, Severin Daiss, Armin Shaukat, Stephan Ritter, Lin Li, Gerhard Rempe},
pdfkeywords = {Schr\"odinger's cat, Schr\"odinger-cat state, optical cat state, cat-state, coherent state superposition, cavity quantum electrodynamics},
colorlinks=true, linkcolor=[RGB]{0,0,0}, citecolor=[RGB]{0,0,0}, filecolor=[RGB]{0,0,0}, urlcolor=[RGB]{0,0,0}}

\newcommand{\bra}[1]{\langle{#1}|}
\newcommand{\ket}[1]{|{#1}\rangle}
\newcommand{\braket}[1]{\langle{#1}\rangle}
\newcommand{\normt}[2]{\ensuremath{\left\langle #1|#2 \right\rangle}}
\newcommand{\up}{{\uparrow}} 
\newcommand{\down}{{\downarrow}} 

\newcommand{\unnumsec}[1]{\refstepcounter{section}\section*{#1}}
\DeclareTextFontCommand{\emph}{\textit}

\makeatletter
\renewcommand{\fnum@figure}{\textbf{Fig.~\thefigure}}
\renewcommand{\@caption@fignum@sep}{~\texorpdfstring{$\boldsymbol{|}$}{$|$}}
\@ifpackageloaded{hyperref}{}{\newcommand{\hyperref}[2][]{#2}}
\@ifpackageloaded{hyperref}{\newcommand{\figref}[1]{\hyperlink{#1}{\ref*{#1}}}}{\newcommand{\figref}[1]{\ref{#1}}}
\makeatother

\begin{document}

\title{Deterministic creation of entangled atom-light Schr{\"o}dinger-cat states}

\author{Bastian~Hacker}
\email[To whom correspondence should be addressed. Email: ]{bastian.hacker@mpq.mpg.de}
\author{Stephan~Welte}
\author{Severin~Daiss}
\author{Armin~Shaukat}
\author{Stephan~Ritter}
\altaffiliation[Present address: ]{TOPTICA Photonics AG, Lochhamer Schlag 19, 82166 Gr{\"a}felfing, Germany}
\author{Lin~Li}
\altaffiliation[Present address: ]{School of Physics, Huazhong University of Science and Technology, Wuhan, China}
\author{Gerhard~Rempe}

\affiliation{Max-Planck-Institut f{\"u}r Quantenoptik, Hans-Kopfermann-Strasse 1, 85748 Garching, Germany}

\begin{abstract}
Quantum physics allows for entanglement between microscopic and macroscopic objects, described by discrete and continuous variables, respectively. As in Schr{\"o}dinger's famous cat gedanken experiment, a box enclosing the objects can keep the entanglement alive. For applications in quantum information processing, however, it is essential to access the objects and manipulate them with suitable quantum tools. Here we reach this goal and deterministically generate entangled light-matter states by reflecting a coherent light pulse with up to four photons on average from an optical cavity containing one atom. The quantum light propagates freely and reaches a remote receiver for quantum state tomography. We produce a plethora of quantum states and observe negative-valued Wigner functions, a characteristic sign of non-classicality. As a first application, we demonstrate a quantum-logic gate between an atom and a light pulse, with the photonic qubit encoded in the phase of the light field.
\end{abstract}

\maketitle

As early as 1935, Schr{\"o}dinger formulated a gedanken experiment \cite{schrodinger1935} with a living cat and a radioactive atom placed inside a box, which is then closed. When the atom decays, it triggers a death mechanism that kills the cat. According to the laws of quantum physics, the decay of the atom occurs at some random time. Consequently, the time of death of the cat is unknown. Mathematically, the situation inside the box is described by an entangled superposition state that is known as a `Schr{\"o}dinger-cat state'. Most remarkable, this state offers a unique access to the atom-cat system, at least in principle. For example, a measurement apparatus that is capable of measuring the atom in a superposition of `not decayed' and `decayed' immediately projects the cat into a superposition of `alive' and `dead'. Such observation thus transfers the superposition state of the microscopic quantum object into the macroscopic classical world, something weird for a cat. In contrast to the entangled Schr{\"o}dinger-cat state, the coherent superposition state of the cat is here denoted as a `cat state'.

While notoriously difficult to create \cite{glancy2008}, several implementations of Schr{\"o}dinger-cat states or just cat states have emerged during recent decades. In all these experiments, coherent states with distinguishable phases mimic the two cat states `dead' and `alive'. Most prominently, Schr{\"o}dinger-cat states were explored using a trapped ion \cite{wineland2013, kienzler2016}, with the vibrational state in the trap taking the role of the cat, and coherent microwave fields confined to superconducting boxes were used in combination with Rydberg atoms \cite{deleglise2008, haroche2013} and superconducting qubits \cite{vlastakis2013}. In the latter experiment, the cat state was also released from the microwave resonator \cite{pfaff2017}.

Modern applications in an open quantum-communication and distributed quantum-networking architecture could benefit from cat states that propagate over some distances. As long as superconducting transmission lines exist for short distances only, optical fields propagating through low-loss optical fibres in the (near) visible part of the electromagnetic spectrum are required. With this backdrop, purely optical realizations of Schr{\"o}dinger-cat states \cite{morin2014, jeong2014, ulanov2017, jeannic2018} as well as approximate cat states \cite{ourjoumtsev2006, ourjoumtsev2007, nielsen2006, takahashi2008, lvovsky2009, namekata2010, gerrits2010} have been demonstrated, but so far only in a heralded fashion. While great effort is under way to release cat states in a quasi-deterministic manner \cite{yoshikawa2013}, our experiment follows a different route. We deterministically create entangled light-matter Schr{\"o}dinger-cat states by reflecting coherent laser pulses from an optical cavity containing a single trapped atom in a controlled superposition of two spin states \cite{wang2005}. We then employ the entanglement to control the flying optical cat state by means of a coherent rotation and subsequent measurement of the atomic spin direction. In principle, this measurement is possible even when the atom and light are far apart.

The experiment reported here integrates the established toolbox of discrete variable qubits, like single photons in a superposition of orthogonal polarization modes, with continuous variable qubits, in our case cat states described by a superposition of coherent states with opposite phases. Such cat states have very promising applications in quantum information processing \cite{ralph2003, gilchrist2004}, mainly because they allow for loss correction \cite{cochrane1999, leghtas2013, bergmann2016} and possibly for fault-tolerant quantum computing \cite{lund2008}. One could therefore envision a hybrid networking architecture where short- to mid-distance communication employs deterministically emitted cat states, and long-distance communication uses probabilistically produced photons in combination with success heralds.

\begin{figure}[t]
\centering
\hypertarget{fig:setup}{}
\includegraphics[width=8.6cm]{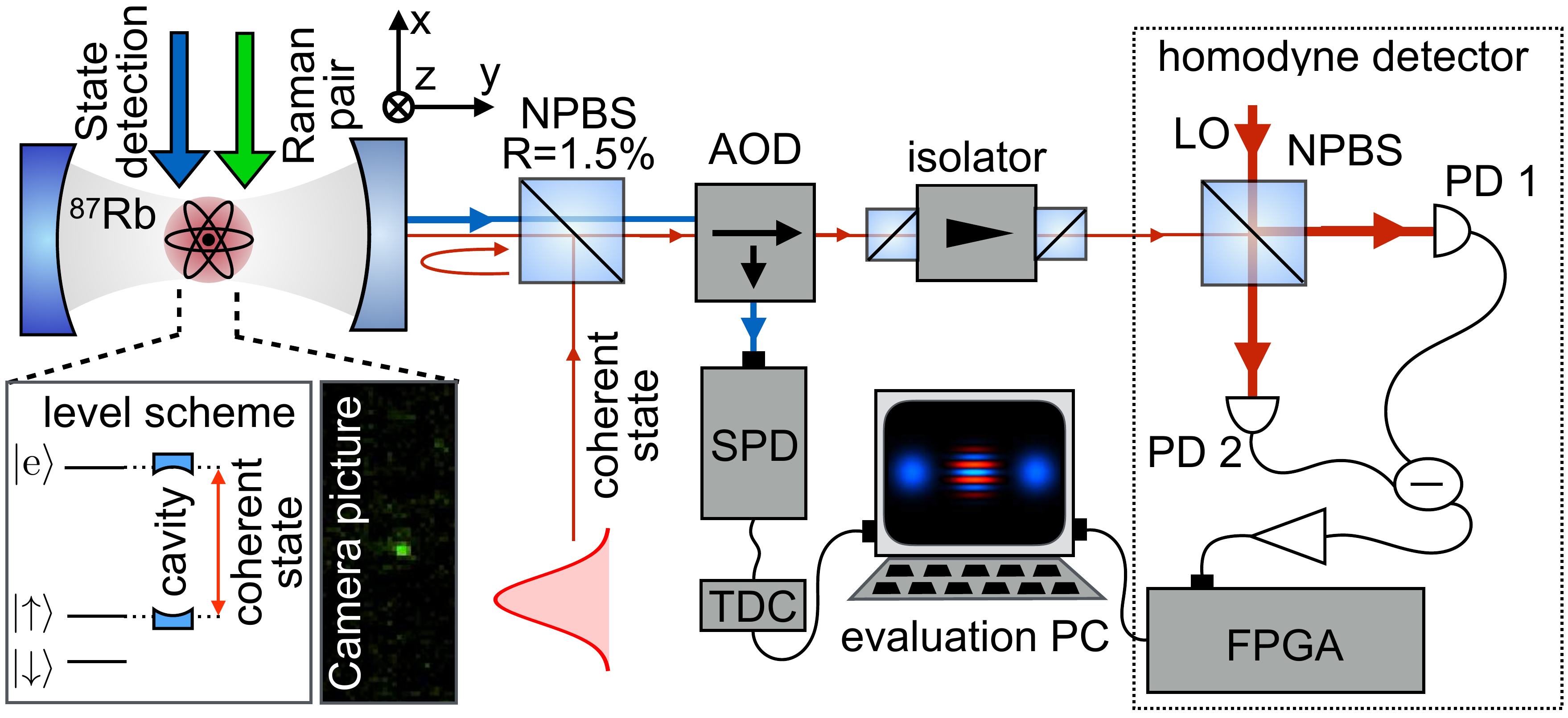}
\caption{\label{fig:setup}
\textbf{Experimental set-up.} A single \textsuperscript{87}Rb atom (imaged via a camera as shown in the inset) is trapped at the centre of a high-finesse cavity. Coherent optical pulses impinge onto the atom-cavity system via the outcoupling mirror. The atom can be modelled as a three-level system as depicted in the level scheme in the lower left inset. After the reflection, the light pulses (full-width at half-maximum of $2.3\,\mathrm{\mu s}$) are directed via an acousto-optical deflector (AOD) through an optical isolator to a homodyne detector with two high-efficiency photodiodes (PDs) to perform quantum state tomography and reconstruct the Wigner function. Fluorescent state detection light is directed to a single photon detector (SPD) to measure the atomic state. The respective clicks are digitized via a time-to-digital converter (TDC) and read out with a PC. A field programmable gate array (FPGA) picks up the homodyne detection signal and can also be read out with the PC.}
\end{figure}

\begin{figure*}[tb]
\hypertarget{fig:protocol}{}
\centering
\includegraphics[width=\textwidth]{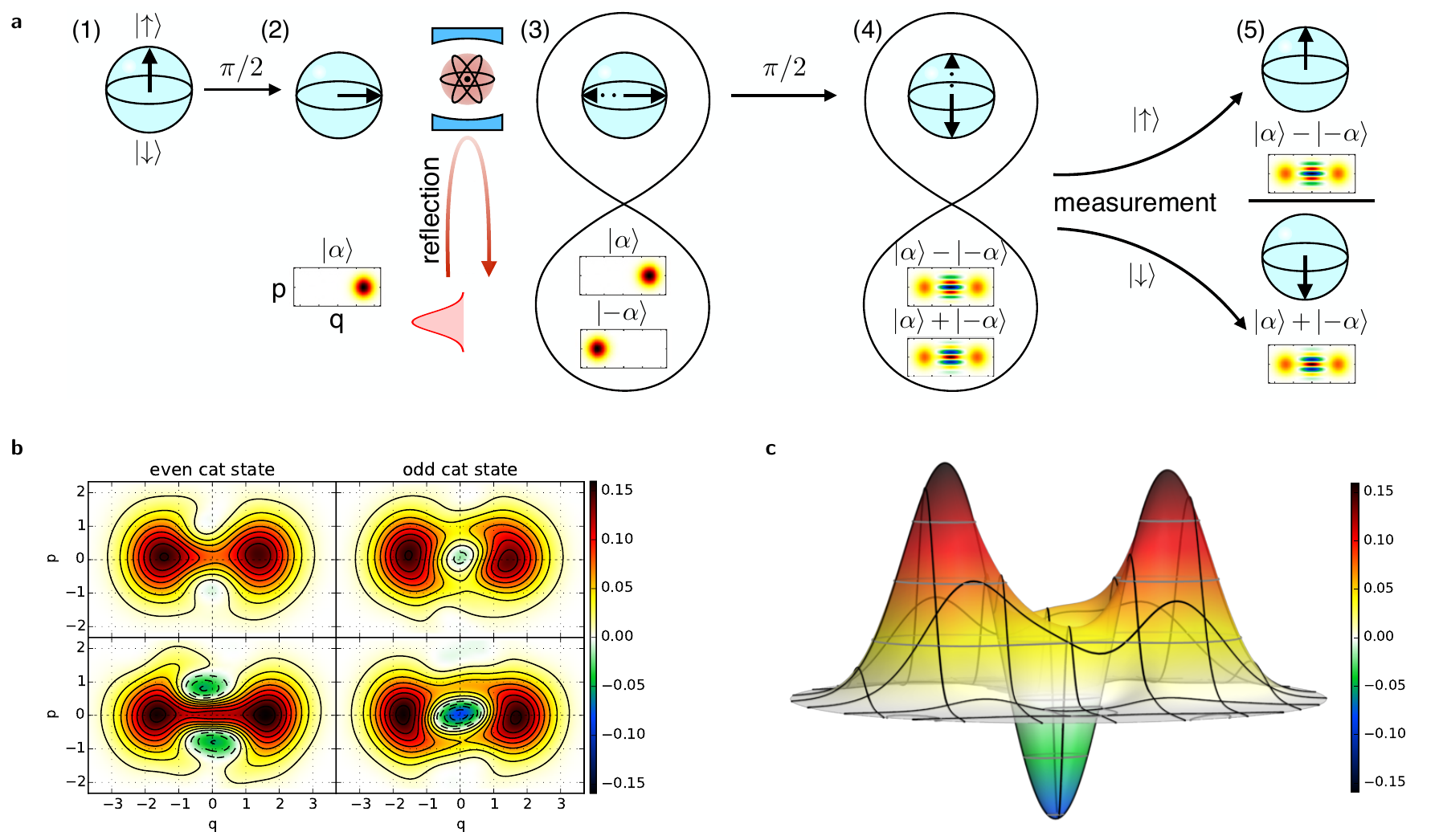}
\caption{
\label{fig:protocol}
\textbf{Cat state generation protocol and Wigner functions of experimentally measured cat states.} \textbf{a}, Steps required to generate a cat state. Initially, the atom is prepared (1) in the state $\ket{\up}$ before a $\pi/2$ rotation is applied (2) to bring it into the coherent superposition $(\ket{\up}+\ket{\down})/\sqrt{2}$. Then, a coherent state $\ket{\alpha}$ is reflected from the cavity and thus an entangled state $(\ket{\up}\ket{\alpha}+\ket{\down}\ket{-\alpha})/\sqrt{2}$ is created (3). A subsequent $\pi/2$ rotation prepares $\frac{1}{2}\bigl[\ket{\up}\bigl(\ket{\alpha}-\ket{-\alpha}\bigr)+\ket{\down}\bigl(\ket{\alpha}+\ket{-\alpha}\bigr)\bigr]$ (4). The last step in the protocol is a state detection on the atom that projects the optical part onto the even or the odd cat state (5).
\textbf{b}, Wigner functions of an even (left) and an odd (right) cat state with $\alpha=1.4$, obtained when the atom is measured in $\ket{\down}$ or $\ket{\up}$, respectively. The fringes between the two Gaussian distributions are shifted by $\pi$ between the two cases. The upper row is reconstructed without loss-correction, whereas the lower row was corrected for $25\%$ propagation and detection losses, which affect the state after its creation at the cavity.
\textbf{c}, Three-dimensional plot of the measured Wigner function of an odd cat state as it emerges from the cavity.}
\end{figure*}

\section*{Experimental set-up}
We employ a single \textsuperscript{87}Rb atom trapped at the centre of a high-finesse ($6{\times}10^5$) optical cavity (Fig.~\figref{fig:setup}; see \hyperref[methods:setup]{Methods}). The atom acts as a three-level system consisting of the two ground states, $\ket{\downarrow}$ and $\ket{\uparrow}$, separated by a microwave transition, and one excited state $\ket{e}$. The atom in state $\ket{\uparrow}$ is strongly coupled to the fundamental mode of the cavity with parameters $(g,\kappa,\gamma)=2\pi{\times}(7.8,2.5,3.0)\,\mathrm{MHz}$. Here, $g$ denotes the atom-photon coupling constant, $\kappa$ is the cavity field decay rate, and $2\gamma$ is the spontaneous atomic decay rate on the transition $\ket{e}\rightarrow\ket{\uparrow}$. The cavity is actively stabilized on this transition with a wavelength of $780\,\mathrm{nm}$. It is furthermore single-sided in the sense that the optical losses $\kappa$ are dominated by the transmission of the mirror through which light is coupled into and out of the cavity with rate $\kappa_r=2\pi{\times}2.3\,\mathrm{MHz}$.

\section*{Protocol}
The protocol \cite{wang2005} to create cat states is outlined in Fig.~\figref{fig:protocol}a. Starting with an atom in the coupling state $\ket{\uparrow}$, we apply a $\pi/2$ pulse to generate an equal superposition state $(\ket{\up}+\ket{\down})/\sqrt{2}$. Next, we reflect a coherent pulse $\ket{\alpha}=e^{-|\alpha|^2/2}\sum_{n=0}^{\infty}(\alpha^n/\sqrt{n!})\ket{n}$ with an arbitrary amplitude $\alpha$ (and the photonic Fock states $\ket{n}$) from the cavity. Here, the mean photon number $\langle{n}\rangle$ is given by $\vert\alpha\vert^2$. This creates a phase shift that depends on the state of the atom \cite{duan2004, reiserer2015}
\begin{align}
\ket{\up}\ket{\alpha}&\rightarrow\ket{\up}\ket{\alpha}, \nonumber\\
\ket{\down}\ket{\alpha}&\rightarrow\ket{\down}\ket{-\alpha}.
\label{eq:phaseshift_eq1}
\end{align}
Thus, after the reflection, the state is $(\ket{\up}\ket{\alpha}+\ket{\down}\ket{-\alpha})/\sqrt{2}$, an entangled state of the atom and the light field. With the initial state of the light field, $\ket{\alpha}$, being classical and potentially macroscopic, the entangled atom-light system is our `Schr{\"o}dinger cat'. A subsequent $\pi/2$ spin rotation on the atom produces the state
\begin{equation}
\label{eq:statefinal}
\frac{1}{2}\bigl[\ket{\up}\bigl(\ket{\alpha}-\ket{-\alpha}\bigr)+\ket{\down}\bigl(\ket{\alpha}+\ket{-\alpha}\bigr)\bigr],
\end{equation}
where the light is in an odd or even superposition of `alive', $\ket{\alpha}$, and `dead', $\ket{-\alpha}$, entangled with the $\ket{\up}$ and $\ket{\down}$ eigenstates of the atom, respectively. Performing a state detection on the atom, the light part of the state (\ref{eq:statefinal}) is projected onto either an odd cat state $(\ket{\alpha}-\ket{-\alpha})/\mathcal{N}^-$ if the atom is observed in $\ket{\up}$ or an even cat state $(\ket{\alpha}+\ket{-\alpha})/\mathcal{N}^+$ if the atom is found in $\ket{\down}$. Here, even and odd refers to the photon number parity of the cat states in Fock space, that is, a state containing only even and odd photon numbers. $\mathcal{N}^\pm$ are the respective normalization factors. Of course, the projection of the entangled atom-light state is inherently random, and thus produces either an odd or an even cat state in each trial. Predictably obtaining one specific state would require feedforward from the atomic spin measurement to the cat state using unitary transformations \cite{jeong2002}. More complex cat states with arbitrary superposition coefficients can be generated by suitable rotations of the atomic state before the measurement. The whole protocol described above lasts $45\,\mathrm{\mu s}$.

The created cat states propagate in a well-defined optical mode and can be sent through an optical fibre that may be part of a larger quantum network. In our experiment, we send the light to a homodyne detector for characterization \cite{lvovsky2009} (see \hyperref[methods:homodyne]{Methods}). Each measurement returns an amplitude of the optical field for a given phase value. The full optical state is defined by its distribution in phase space, the Wigner function $W(q,p)$ that we reconstruct from a larger set of measurements. Here, $q$ and $p$ are the field quadratures spanning phase space. Our reconstruction basis is the density matrix $\rho$ in truncated Fock space.

\section*{Results}
We observe the characteristic Wigner functions (Fig.~\figref{fig:protocol}b,c) with two Gaussian peaks and interference fringes in the centre that encode the coherent nature of the superposition state \cite{schleich1991}. The even cat state displays a local maximum and the odd cat state a local minimum at the centre of the Wigner distribution. The fidelities of these measured states with ideal even and odd cat states are $59.2(6)\%$ and $51.2(6)\%$, respectively. Applying a correction \cite{dariano1995} for $25\%$ combined propagation and detection loss, we obtain $68.0(9)\%$ and $62.8(8)\%$, respectively.

\begin{figure}[tb]
\centering
\hypertarget{fig:visibilityplot}{}
\includegraphics[width=8.6cm]{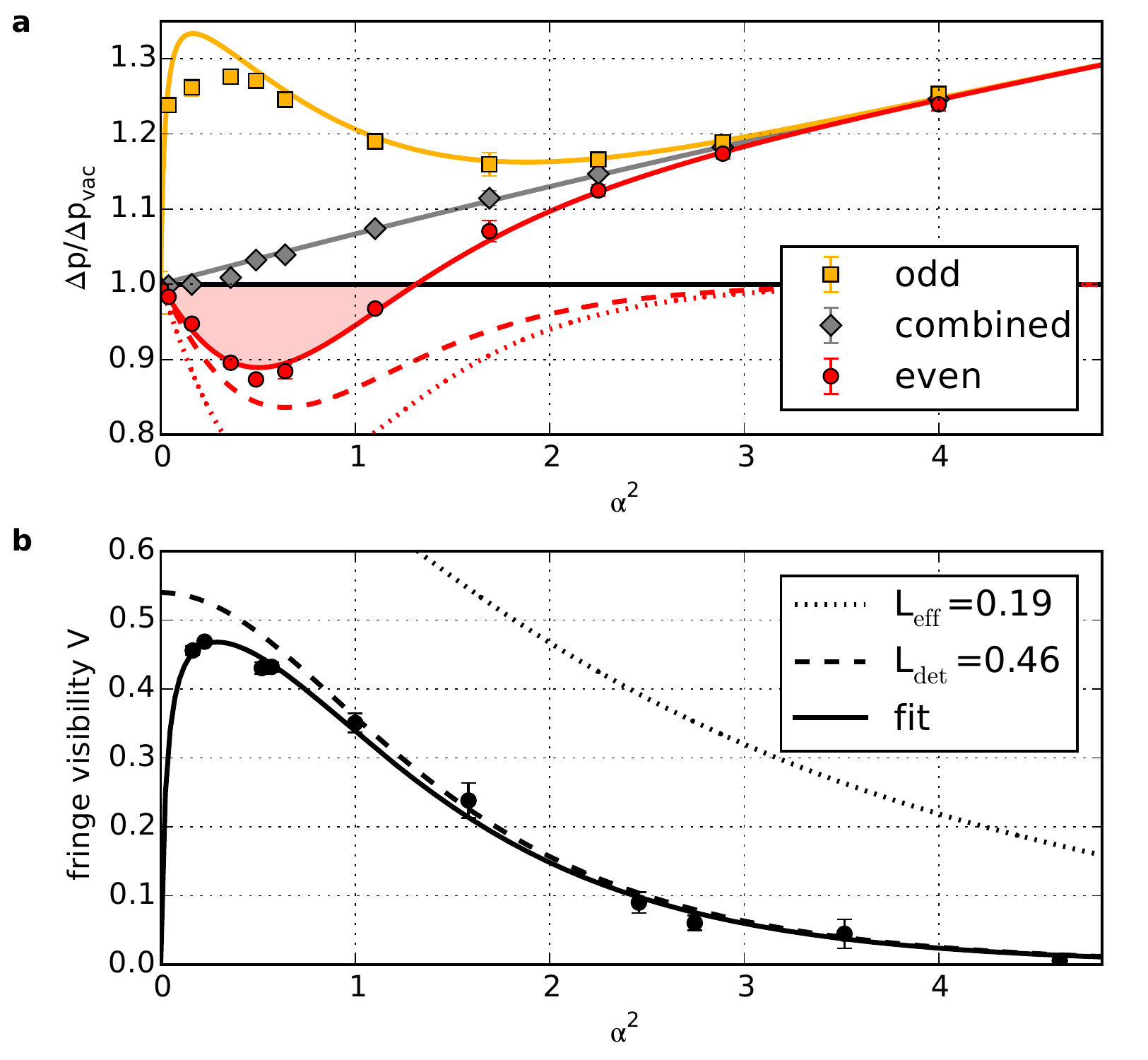}
\caption{\label{fig:visibilityplot}
\textbf{Non-classical properties of measured cat states.}
\textbf{a},~Squeezing. Even cat states are squeezed with reduced $p$-quadrature fluctuations for low $\alpha^2$ (light red area), with a maximum of $1.18(3)\,\mathrm{dB}$. Dotted and dashed lines show the theoretical prediction for even cat states with losses $L_\textnormal{eff}=0.19$ and $L_\textnormal{det}=0.46$ for the states immediately after the cavity and at the homodyne set-up, respectively. The full model (solid lines) for even (red), odd (yellow) and combined (grey) cat states is affected by atomic state detection errors of $1.3\%$ at small $\alpha^2$ and random noise in the phase angle $\phi$ at larger $\alpha^2$. The latter limits the observable squeezing to states with $\alpha^2<1.3$. \textbf{b},~Interference fringe visibility $V$ of measured Wigner functions. The dotted and dashed lines show the effect of experimental losses $L_\textnormal{eff}$ and $L_\textnormal{det}$, respectively, which reduce the visibility with increasing $\alpha^2$. For $\alpha^2$ close to 0, errors in the atomic state detection dominate and cause the visibility to drop (solid line, fitted to data). All data points are without correction for optical losses. Error bars depict statistical standard errors.}
\end{figure}
A characteristic feature of cat states, which has no classical explanation, is the existence of negative regions in the Wigner function. We observe a minimum value of $-0.016(4)$ for the odd cat state with $\alpha=1.4$, as shown in Fig.~\figref{fig:protocol}b. If we correct for propagation and detection losses, the minimum value becomes $-0.083(5)$ (Fig.~\figref{fig:protocol}b,c). An intrinsic property of even cat states is `squeezing', that is, the uncertainty of one of the field quadratures, $\Delta p$, is smaller than for a coherent state \cite{buzek1992}. Experimental results are displayed in Fig.~\figref{fig:visibilityplot}a that shows the minimal width of the measured quadrature distributions as a function of $\alpha^2$. We observe squeezing with a maximal amount of $1.18(3)\,\mathrm{dB}$ at $\alpha^2=0.5$ (for details see Supplementary Section \ref{supplement:squeezing}). The measured values follow the theoretical prediction $\Delta p/\Delta p_\textnormal{vac}=\sqrt{1-4(1-L)\alpha^2/(1+\exp{2\alpha^2})}$ with relative optical losses $L$ and the uncertainty of the vacuum state $\Delta p_\textnormal{vac}=1/\sqrt{2}$. With increasing $\alpha^2$ the width $\Delta p$ increases due to experimental noise in the optical phase $\phi$ (angle between the two Gaussian distributions in phase space) of $\Delta \phi=0.06\pi$, attributed to residual fluctuations of the cavity and laser frequencies of $0.2\,\mathrm{MHz}$.
\par\bigskip

\noindent\textbf{Losses.}
It is now worthwhile to discuss the optical losses. These losses reduce the visibility $V$ of the non-classical coherences \cite{spagnolo2009} which appear as fringes in the Wigner function. To quantify this effect, we express the visibility as the difference between even- and odd-cat Wigner functions at the centre of the cat state, $V=\frac\pi2(W_\textnormal{even}(0,0)-W_\textnormal{odd}(0,0))$. The theoretical calculation (see \hyperref[methods:losses]{Methods}) gives $V=\sinh(2(1-L)\alpha^2)/\sinh(2\alpha^2)$ for an initially ideal cat state. Measured visibilities are presented in Fig.~\figref{fig:visibilityplot}b.

In our experiment, losses stem from the finite cooperativity of the atom-cavity system, $C=g^2/(2\kappa\gamma)=4.1$, and the reduced escape efficiency $\eta_\textnormal{esc}=\kappa_r/\kappa=0.92$ of the cavity, which does not reflect all of the incoming light. For a coherent input of size $\alpha$ the effective size of the output cat state, $\alpha_\textnormal{out}=|\alpha_+-\alpha_-|/2$, that is the phase-space distance between the two coherent contributions, is obtained from input-output theory (see \hyperref[methods:losses]{Methods}) as $\alpha_\textnormal{out}=\eta\alpha$ with
\begin{equation}
\eta
= \frac{\kappa_r}{\kappa}\frac{g^2}{g^2+\kappa\gamma}
= \eta_\textnormal{esc}\frac{C}{C+1/2}
= 0.81.
\end{equation}
This corresponds to a loss of $L_\textnormal{cav}=1-\eta^2=34\%$.

Remarkably, not all these losses reduce the coherence of the created cat state like a classical absorber would do. This is caused by some losses that effectively occur before the cat state is created. Theoretically, we find that the effective coherence-reducing losses $L_\textnormal{eff}$ are (see \hyperref[methods:losses]{Methods})
\begin{equation}
\label{eq:Leff}
L_\textnormal{eff} = 1-\eta = 19\%,
\end{equation}
(much less than $L_\textnormal{cav}$). Further losses $L_\textnormal{p+d}$ occur during propagation and detection of the cat state. These are mainly optical absorption losses $L_\textnormal{o}=14\%$. A small mode mismatch between the signal and the local oscillator used in the homodyne set-up can be treated as an additional loss $L_\textnormal{m}=6.0\%$. The detector there comprises two photodiodes with $L_\textnormal{d}=1.5\%$. The subtracted photodiode signals are fed into a field-programmable gate array (FPGA) for further processing. We identified several more noise sources that add effective losses of $L_\textnormal{n}=5.5\%$ in total (Supplementary Section \ref{supplement:loss_budget}). Combining these loss channels according to $1-L_\textnormal{p+d}=\prod_i(1-L_{i})$ with $i\in\{\textnormal{o}, \textnormal{m}, \textnormal{d}, \textnormal{n}\}$, we obtain $L_\textnormal{p+d}=25\%$. The fringe visibilities in our reconstructed states (solid line in Fig.~\figref{fig:visibilityplot}b) match the theoretical model with combined total losses $L_\textnormal{det}=46(1)\%$. In addition to the losses $L_\textnormal{eff}$ and $L_\textnormal{p+d}$, the observed visibility suffers from an imperfect atomic state detection as well as laser or cavity phase fluctuations. Yet $L_\textnormal{det}$ is well below the threshold of $50\%$, where the Wigner function becomes entirely positive \cite{spagnolo2009}. While the protocol puts no intrinsic constraint on the size of produced cat states, in our experiment the size is limited by these losses to $\alpha^2\lesssim4$.
\par\bigskip

\noindent\textbf{Generalized cats.}
\begin{figure}[tb]
\centering
\hypertarget{fig:scan}{}
\includegraphics[width=\columnwidth]{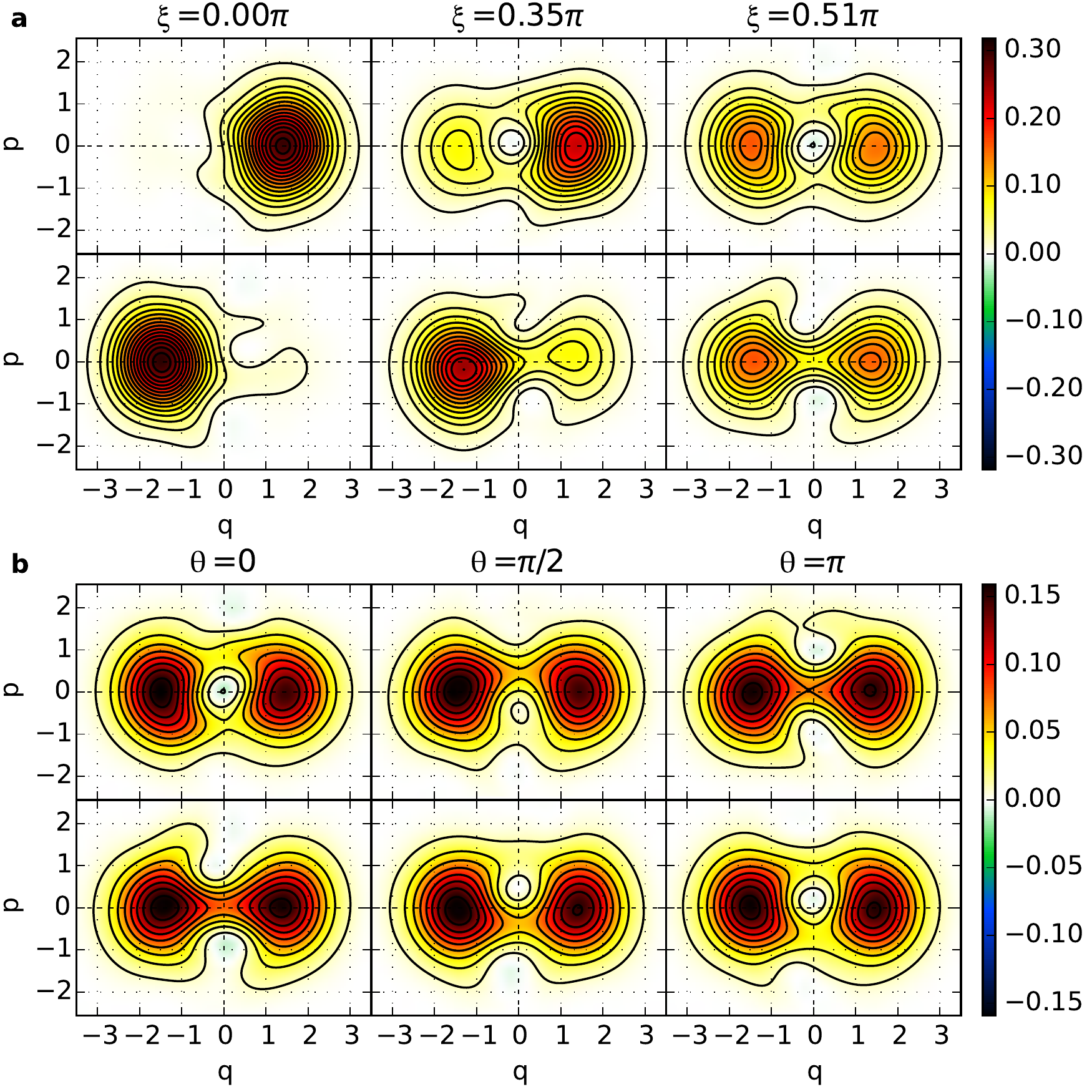}
\caption{
\label{fig:scan}
\textbf{Control over cat state degrees of freedom.}
\textbf{a}, Measured Wigner functions of optical states for different rotation angles $\xi$ of the atom. The top row shows the optical states after post-selecting on a measurement outcome $\ket{\up}$, while the bottom row shows the complementary data when the atom was in $\ket{\down}$ (as in \textbf{b}). A continuous transition from the coherent states $\ket{\alpha}$ and $\ket{-\alpha}$ to the odd and even cat states, respectively, is observable. In this measurement, we chose $\theta=0$, $\phi=\pi$ and $\alpha=1.4$.
\textbf{b}, The phase $\theta$ for cat states $(\ket{\alpha}\mp e^{i\theta}\ket{-\alpha})/\mathcal{N}$ is controlled by the rotation axis of the atomic state. We observe a continuous transition from an odd into an even cat state and vice versa. No loss-correction was applied. Data with more intermediate steps are shown in Supplementary Section \ref{supplement:control-cat}.
}
\end{figure}
The even and odd cat states shown so far are special cases of more general coherent state superpositions of the form
\begin{equation}
\ket{\psi_{\textnormal{cat}}} = \frac{1}{\mathcal{N}} \left(\cos(\xi/2)\ket{\alpha} + e^{i\theta}\sin(\xi/2)\ket{e^{i\phi}\alpha}\right)
\label{eq:catparameters}
\end{equation}
with parameters $\alpha$, $\phi$, $\theta$ and $\xi$. Our experiment allows the control of all these parameters, the phase and modulus of $\alpha$, the optical phase $\phi$ between the coherent contributions, the superposition phase $\theta$ that determines even or odd cat states and the population fraction of the two coherent contributions $\xi$. This opens the possibility to create more complex states that could be useful for continuous-variable error correction codes \cite{cochrane1999, leghtas2013, bergmann2016}.

Specifically, the angle of the last spin rotation (step 4 in Fig.~\figref{fig:protocol}a) controls the relative amplitude of the two coherent state contributions in the generated cat state. Scanning the Raman-pulse area $\xi$ from $0$ to $\pi/2$ continuously transforms the optical state (after measuring the atom) from a coherent state into a cat state with equal amplitudes of coherent contributions (Fig.~\figref{fig:scan}a).

The cat state in equation~(\ref{eq:catparameters}) can be continuously tuned from an even cat state into an odd cat state via a change of the phase $\theta$ of the final $\pi/2$ rotation (step 4 in Fig.~\figref{fig:protocol}a). This phase is imprinted onto the observed interference fringes in the Wigner function. Figure~\figref{fig:scan}b shows the continuous transition from an even into an odd cat state and vice versa for $0\leq\theta\leq\pi$. The optical phase $\phi$ is varied between $0$ and $2\pi$ via a detuning between the impinging light and the cavity resonance. On resonance, $\phi=\pi$. Data showing control over $\alpha$, $\phi$, $\theta$ and $\xi$ in more detail are presented in Supplementary Section \ref{supplement:control-cat}.
\par\bigskip

\begin{figure}[tb]
\centering
\hypertarget{fig:stokes}{}
\includegraphics[width=8.6cm]{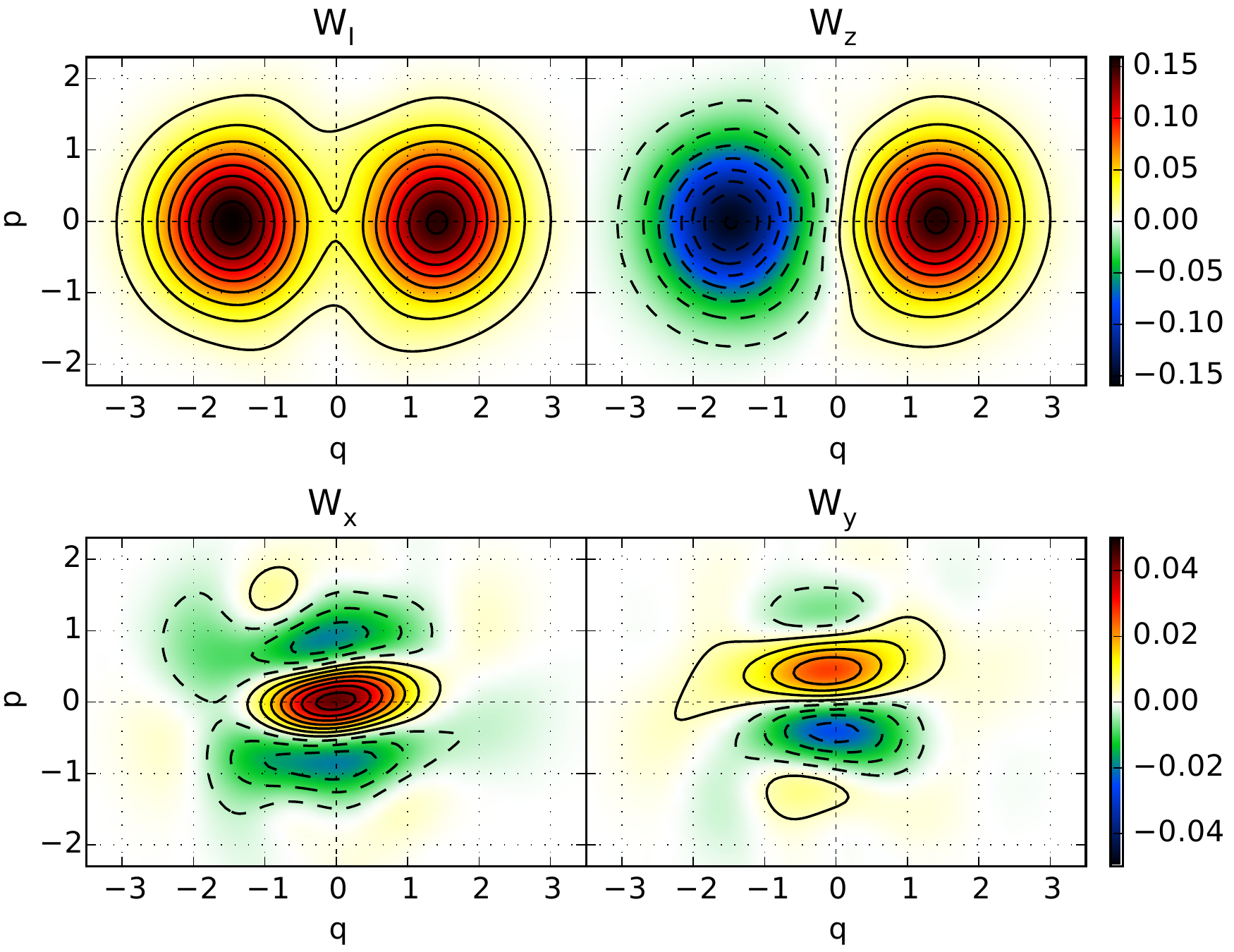}
\caption{\label{fig:stokes}
\textbf{Entanglement between atom and cat state.}
The combined atom-cat density matrix is represented through four optical joint Wigner functions $W_i$ obtained for atomic spin measurements in the respective bases. $W_I$ is the optical state when the atom is not evaluated. $W_z$ is the difference of the two contributing coherent peaks. $W_x$ and $W_y$ are the interference fringes of cat states with two different phases $\theta=0$ and $\theta=-\pi/2$, respectively. Here, $\alpha=1.4$ and no loss-correction was applied.}
\end{figure}

\noindent\textbf{Entanglement.}
So far we have used the light-matter entanglement provided by the Schr{\"o}dinger-cat state to produce an optical cat state by making a measurement on the atom. We now show that light and matter are indeed entangled by observing an entanglement witness. To this end, we measure the atomic state in three different detection bases via the application of spin rotation pulses. Representing the spin in terms of Pauli matrices $\sigma_i=\{\sigma_I, \sigma_x, \sigma_y, \sigma_z\}$ and the corresponding optical state as joint Wigner functions \cite{vlastakis2015} $W_i=\{W_I, W_x, W_y, W_z\}$, which play the role of Stokes parameters, allows us to express the full density matrix of the entire atom-cat system as $\rho_{ac}=\frac{1}{2}\sum_{i}\sigma_i\otimes W_i$. Here $W_I$ is the average photonic density matrix with the atom traced out. $W_{x,y,z}$ are sums of photonic density matrices weighted with the outcome of the atomic spin measurement of $\pm1$ for outcomes $\ket{\up}$ and $\ket{\down}$, respectively, in the measurement bases $x$, $y$ and $z$. Wigner representations of our measured $W_i$ are displayed in Fig.~\figref{fig:stokes}. The obtained density matrix $\rho_{ac}$ has a negativity \cite{vidal2002} $N=0.057(5)$ for an input state with $\alpha=1.4$. The statistically significant positive value witnesses the entanglement.

\begin{figure}[tb]
\centering
\hypertarget{fig:atom-cat-gate-truthtable}{}
\includegraphics[width=8.6cm]{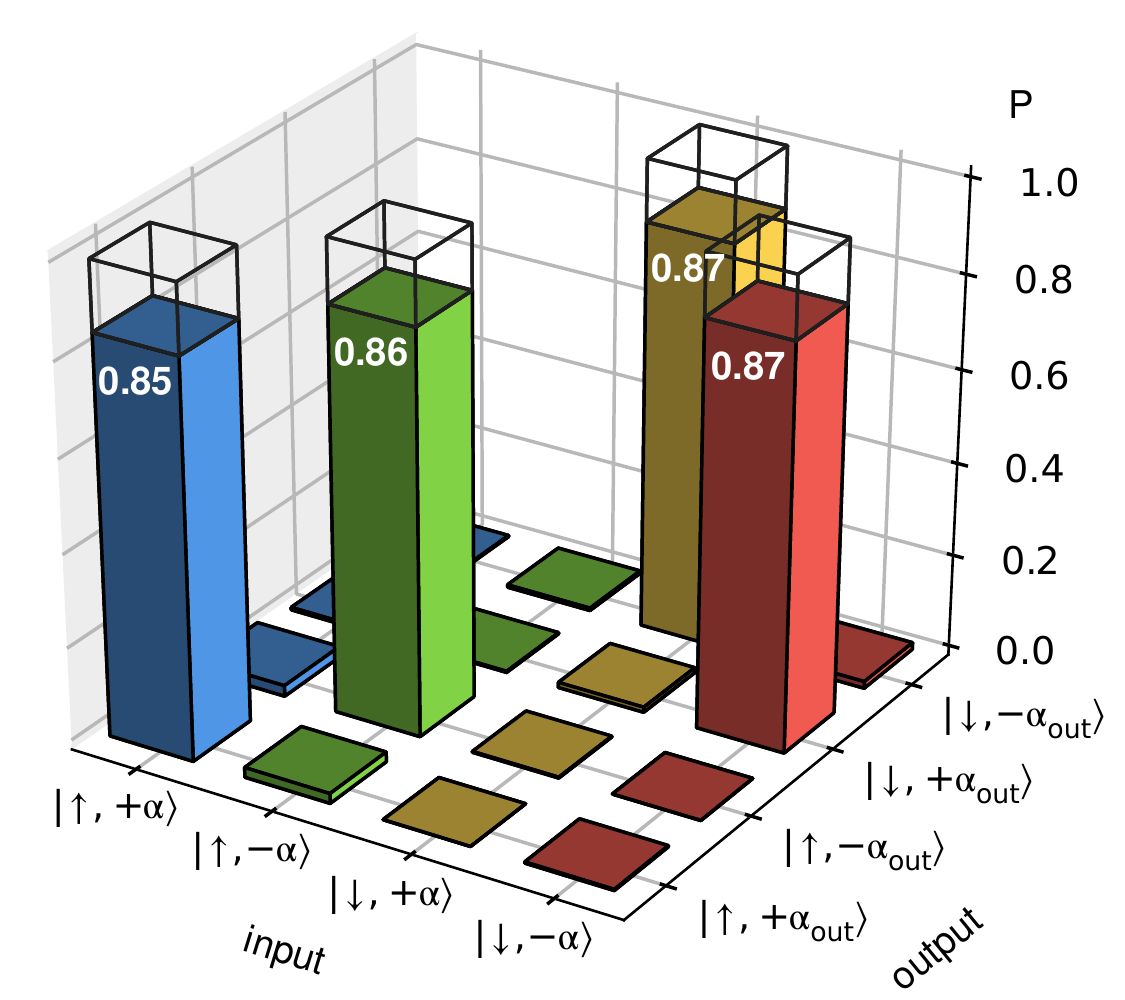}
\caption{\label{fig:atom-cat-gate-truthtable}
\textbf{Truth table of the atom-cat gate.}
For the choice of basis states $\ket{\up}/\ket{\down}\otimes \ket{\alpha}/\ket{-\alpha}$, the gate operates as a CNOT where the atom controls the optical target qubit. With the atom in $\ket{\up}$, the optical state is unaffected (bars on the diagonal), whereas the atom in $\ket{\down}$ causes the coherent state to flip phase (off-diagonal bars) as predicted by equation~(\ref{eq:phaseshift_eq1}). The open bars show an ideal CNOT. The optical input states were coherent with $\alpha=\pm1.4$. The overlap with the expected output states with $\alpha_\textnormal{det}=1.0$ is calculated as the overlap between expected and measured states. Imperfections are mainly due to broadened phase-space distributions.}
\end{figure}
\par\bigskip

\noindent\textbf{Quantum gate.}
The reflection mechanism (equation~(\ref{eq:phaseshift_eq1})) that has so far been used to create entanglement is essentially a quantum-logic gate between an atom and a light field. Here, the coherent states $\ket{\alpha}$ and $\ket{-\alpha}$ form an orthogonal qubit basis as long as $\alpha$ is large enough. For our value, $\alpha=1.4$, the overlap of these basis states is $|\braket{-\alpha|\alpha}|^2=\exp(-4|\alpha|^2)=3.9{\times}10^{-4}$, small enough for a good qubit. To probe our mechanism as a gate, we employ a basis set of input states $\ket{\up,\alpha},\ket{\up,-\alpha},\ket{\down,\alpha},\ket{\down,-\alpha}$. In this basis, the gate acts as a CNOT \cite{nielsen2000} with the truth table
\begin{align}
\ket{\up,+\alpha}&\rightarrow\ket{\up,+\alpha}\nonumber\\ 
\ket{\up,-\alpha}&\rightarrow\ket{\up,-\alpha}\nonumber\\
\ket{\down,+\alpha}&\rightarrow\ket{\down,-\alpha}\nonumber\\
\ket{\down,-\alpha}&\rightarrow\ket{\down,+\alpha}.
\end{align}
Here the atom serves as the control qubit that can flip the optical target qubit. We characterize this gate with coherent pulses containing a mean photon number of $\vert\alpha\vert^2=\langle{n}\rangle\approx2$, the same value as in the measurements confirming the entanglement between the atom and the light field. The quantum nature of the gate has thus been demonstrated in the context of Fig.~\figref{fig:stokes}. The classical truth table is shown in Fig.~\figref{fig:atom-cat-gate-truthtable} and exhibits the expected CNOT behaviour. Here, the optical losses change the input states $\ket{\pm\alpha}$ to output states $\ket{\pm\alpha_\textnormal{det}}$ with $\alpha_{\textnormal{det}}=\alpha\sqrt{1-L_{\textnormal{det}}}=1.0$. This change is well-characterized and thus predictable, and the overlap between the two output states is still small ($1.8{\times}10^{-2}$). We calculate the mean fidelity of the observed output states with the expected output states and find values around $86\%$. The reduction from $100\%$ comes mainly from phase noise due to cavity and laser frequency fluctuations that broaden the coherent-state Wigner function of the reflected field in the azimuthal direction. If we discriminate the coherent state qubits simply between negative and positive $q$-quadratures, the fidelity of the truth table becomes $96\%$. Note that the gate acts as a CPHASE in the $\ket{\up}/\ket{\down}\otimes \ket{\textnormal{cat}_\pm}$ basis.

\section*{Conclusion}
The experiment has demonstrated unprecedented control over all relevant parameters of optical cat states, making them a resource for continuous-variable quantum information processing with coherent-state superpositions as qubits \cite{ralph2003, gilchrist2004}. The useful size of these qubits depends on the optical losses, elimination of which is the key challenge for future improvements. Optical losses in the cavity (equation~(\ref{eq:Leff})) can be reduced with improved cavity parameters. For example, parasitic losses $\kappa-\kappa_r$ could be lowered by reducing surface scattering and absorption. Likewise, the atom-cavity coupling rate $g$ could be increased with a smaller mode volume, either by reducing the mirror distance or by decreasing the mirror radius of curvature. Here, microscopic fibre cavities offer improvements by more than an order of magnitude. Propagation losses can be mitigated by loss-correction codes that are possibly similar to those that have already been tested in the microwave domain \cite{ofek2016}. Optical cats as qubits could therefore be a promising alternative to single photons for quantum communication in a future quantum internet \cite{nielsen2006, kimble2008}, at least when losses are not too severe.

It should be emphasized that in contrast to all protocols so far realized with single optical photons as qubits \cite{reiserer2015}, the creation and detection of Schr{\"o}dinger-cat states as implemented here prepares and verifies an entangled light-matter state in each trial, without any post-se\-lec\-tion.  One could utilize this entanglement in a hybrid Bell test \cite{teo2013, kwon2013, vlastakis2015} that takes advantage of the deterministic nature of the entangling operation as well as the near-unity detection efficiency for both the atomic and the photonic state. Our atomic qubit could also be mapped to the polarization of a single photon through the readout process reported in ref.~\cite{kalb2015}. This would extend our atom-light entanglement towards purely optical photon-light entanglement \cite{andersen2015}, a hybrid entanglement between discrete and continuous variables states of light.

\vspace{\baselineskip}
This work:
\newenvironment{extrabibitem}{\list{}{\leftmargin=1em\rightmargin=0em}\item[]}{\endlist}
\begin{extrabibitem}
\small
Hacker, B., Welte, S., Daiss, S., Shaukat, A., Ritter, S., Li, L. \& Rempe, G.
Deterministic creation of entangled atom-light Schr{\"o}dinger-cat states.
\href{https://www.doi.org/10.1038/s41566-018-0339-5}{\textit{Nat.\ Photon.} \textbf{13}, 110--115} (2019).
\end{extrabibitem}

\vspace{2\baselineskip}
\noindent\textbf{\large Acknowledgments}\\
The authors thank J.I.\ Cirac, S.\ D{\"u}rr and O.\ Morin for valuable ideas and discussions. This work was supported by the Deutsche Forschungsgemeinschaft via the excellence cluster Nanosystems Initiative Munich (NIM) and the EU flagship project Quantum Internet Alliance (QIA). S.W.\ was supported by Elitenetzwerk Bayern (ENB) through the doctoral program Exploring Quantum Matter (ExQM).
\vspace{2\baselineskip}

\noindent\textbf{\large Author Contributions}\\
Experimental data were taken and analysed by B.H., S.W., S.D.\ and L.L.. The homodyne detection set-up was built by B.H., S.W., S.D., A.S., S.R.\ and L.L.. The paper was written by B.H., S.W.\ and G.R., with input from all authors.

\clearpage

\unnumsec{Methods}

\noindent\textbf{\phantomsection\label{methods:setup}Experimental set-up.}
A schematic of the set-up is shown in Fig.~\figref{fig:setup}.
We used an optical Fabry-P{\'e}rot cavity with a mirror separation of $0.5\,\mathrm{mm}$ and finesse of $F=6{\times}10^4$. The relevant cavity quantum electrodynamics parameters were $(g,\kappa,\gamma)=2\pi(7.8,2.5,3.0)\,\mathrm{MHz}$, placing our atom-cavity system in the strong coupling regime. In our notation, $g$ is the atom-photon coupling rate, $\kappa$ the cavity field decay rate and $2\gamma$ the spontaneous decay rate of the excited atomic state $\ket{e}$. With this set-up, the atoms are initially trapped in a magneto-optical trap and transferred into the cavity. Once an atom arrives at the centre of the cavity, a three-dimensional standing-wave optical lattice is switched on. We used a red-detuned dipole trap ($1064\,\mathrm{nm}$) perpendicular to the cavity axis along the $x$ direction in Fig.~\figref{fig:setup} and two blue-detuned ($771\,\mathrm{nm}$) repulsive standing waves in the $y$ and $z$ directions. The blue-detuned traps confine the atom to their standing wave nodes, where the intensity-dependent (and therefore fluctuating) light shift is small. The atom is held in this configuration on average for $10\,\mathrm{s}$. We monitored the atom with a camera and a high-numerical-aperture ($\textnormal{NA}=0.4$) objective. Conditioned on the presence of an atom in the cavity, we started the experimental protocol. Initially, the atom is pumped to the state $\ket{\up}=\ket{5^2S_{1/2}, F{=}2,m_F{=}2}$ via a right-circularly polarized pump laser along the cavity axis on the $F{=}2\leftrightarrow F'{=}3$ transition. The pump light is applied for $120\,\mathrm{\mu s}$, and successful pumping of the atom is heralded by a reduction of the cavity transmission \cite{thompson1992}. After this step, the atom can effectively be considered as a three-level system consisting of the states $\ket{\up}$, $\ket{\down}=\ket{5^2S_{1/2}, F{=}1,m_F{=}1}$ and $\ket{e}=\ket{5^2P_{3/2}, F'{=}3,m_F{=}3}$. Spin rotations and the generation of superposition states of $\ket{\uparrow}$ and $\ket{\downarrow}$ are realized with a pair of Raman lasers (green arrow in Fig.~\figref{fig:setup}). The frequency difference between these two orthogonally polarized laser beams corresponds to the frequency difference between the states $\ket{\downarrow}$ and $\ket{\uparrow}$ of $6.8\,\mathrm{GHz}$. The Raman lasers are $131\,\mathrm{GHz}$ red-detuned from the \textsuperscript{87}Rb D1 transition at $795\,\mathrm{nm}$ and impinge onto the atoms from the side. Each application of the cat state protocol is followed by Sisyphus cooling, which is applied to the atom for $1700\,\mathrm{\mu s}$. The entire protocol is repeated at a rate of $500\,\mathrm{Hz}$.
\par\bigskip

\noindent\textbf{\phantomsection\label{methods:homodyne}Homodyne detection.}
We characterized the optical states via balanced homodyne detection\cite{lvovsky2009}. In this process, the signal is mixed with a $1.8\,\mathrm{mW}$ local oscillator (LO) beam of a continuous-wave laser on a 50/50 non-polarizing beamsplitter (NPBS). The intensities of the two output beams are measured on separate photodiodes (Hamamatsu S3883 with glass windows removed) and subtracted electronically. This difference signal is proportional to the signal beam amplitude and amplified to a measurable level. We recorded a time trace of each signal pulse with a field-programmable gate array (FPGA), and averaged the amplitude over the known temporal mode of the pulse. This results in one electric field amplitude (quadrature) in each homodyne measurement. The quadrature is a projection of the amplitude in phase space under a projection angle that is given by the relative phase of the LO with respect to the signal beam. We do not interferometrically stabilize this phase in our homodyne measurements. The phase thus changes due to thermal drifts of the employed optics and a $1\,\mathrm{Hz}$ detuning of the LO that ensures uniform sampling over all projection angles. For each homodyne measurement the relative phase between signal and LO is determined separately. To this end, the power of the reflected beam is increased by a factor of $30$ for a sufficient signal-to-noise ratio and the LO phase is swept for $100\,\mathrm{\mu s}$ over $2\pi$ to record a full interference oscillation. In order not to lose the atom from the intra-cavity dipole traps during this measurement, it is pumped to the non-coupling state $\ket{\down}$, which is almost completely decoupled from the cavity. In each shot, the phase of the reflected beam with respect to the local oscillator can be determined with an accuracy of $2^\circ$. Phase drifts within a few $100\,\mathrm{\mu s}$ between the experiment and the phase determination are negligible.

For state reconstruction, we performed many repeated measurements, each resulting in one quadrature value and a corresponding LO phase that determines the projection angle in phase space. We acquire on the order of $10^4$ samples for each measured optical state, and each outcome can be considered as one projective measurement of the density matrix $\rho$. The density matrix in truncated Fock space is obtained via maximum likelihood estimation using the iterative $R\rho R$ algorithm \cite{lvovsky2004}. This ensures a physical density matrix and allows for the correction of optical losses. We obtain an estimate for the uncertainty (covariance matrix) of each reconstructed density matrix using the Hessian of the likelihood function \cite{banaszek1999}. The density matrix fully characterizes the quantum state and therefore allows us to compute all derived quantities, such as the Wigner function, photon statistics or fringe visibility. Some quantities, such as squeezing, can directly be derived from the raw data and do not depend on the density matrix reconstruction.
\par\bigskip

\noindent\textbf{\phantomsection\label{methods:losses}Analytic treatment of cavity losses.}
Our atom-cavity system is well described by input-output theory\cite{kuhn2015} in the limit of slowly varying light intensities and low atomic excitation probability. For simplicity, we assume the case where cavity, atom and coherent input light $\ket{\alpha}$ are on resonance, and all amplitudes are real. We consider four modes: cavity reflection $\ket{r}$, cavity transmission $\ket{t}$, mirror losses $\ket{m}$ and scattering via the atom $\ket{a}$. The modes can be considered coherent with amplitudes
\begin{eqnarray}
r_{\down/\up} &=& \frac{Ng^2+(\kappa-2\kappa_r)\gamma}{Ng^2+\kappa\gamma}\,\alpha
\\
t_{\down/\up} &=& \frac{2\sqrt{\kappa_r\kappa_t}\gamma}{Ng^2+\kappa\gamma}\,\alpha
\\
m_{\down/\up} &=& \frac{2\sqrt{\kappa_r\kappa_m}\gamma}{Ng^2+\kappa\gamma}\,\alpha
\\
a_{\down/\up} &=& \frac{2\sqrt{\kappa_r\gamma}\sqrt{N}g}{Ng^2+\kappa\gamma}\,\alpha
,
\end{eqnarray}
respectively. Here, $N$ is the number of coupling atoms, $N=0$ for $\ket{\down}$ and $N=1$ for $\ket{\up}$, $\kappa_i$ are optical field decay rates into the respective modes and $\kappa=\kappa_r+\kappa_t+\kappa_m$.

With a coherent input $\ket{\alpha}$ of amplitude $\alpha$, the output will be a superposition of two coherent fields of amplitude $r_{\down}$ and $r_{\up}$, which we call $\ket{r_\down}$ and $\ket{r_\up}$, respectively. If the moduli of the two amplitudes differ, the resulting cat state in phase space will be off-centred, which is easily correctable with a displacement. In our experiment however, the two amplitudes are nearly identical. The total size of the cat state is given by the peak separation
\begin{equation}
\alpha_{\textnormal{out}} = \frac{1}{2}\vert r_{\up}-r_{\down}\vert = \frac{\kappa_r}{\kappa}\frac{g^2}{g^2+\kappa\gamma}\alpha = \eta\alpha\ ,
\end{equation}
where we use the definition
\begin{equation}
\eta:=\frac{\kappa_r}{\kappa}\frac{g^2}{g^2+\kappa\gamma}\ (=0.81)
\ .
\end{equation}
The total losses can be expressed as $L_\textnormal{cav}=1-\alpha_{\textnormal{out}}^2/\alpha^2=1-\eta^2\ (=0.34)$.

Henceforth we consider the reflected light modes $\ket{r_{\up}}$, $\ket{r_{\down}}$ and the loss modes $\ket{l_{\up}}:=\ket{t_{\up}}\ket{m_{\up}}\ket{a_{\up}}$ and $\ket{l_{\down}}:=\ket{t_{\down}}\ket{m_{\down}}\ket{a_{\down}}$, which have overlaps of
\begin{equation}
\normt{r_\up}{r_\down} = e^{-2\eta^2\alpha^2}\ ,
\end{equation}
\begin{equation}
\normt{l_\up}{l_\down} = \normt{t_\up}{t_\down}\normt{m_\up}{m_\down}\normt{a_\up}{a_\down} = e^{-2(1-\eta)\eta\alpha^2}\ ,
\end{equation}
\begin{equation}
\bra{l_\up}\normt{r_\up}{r_\down}\ket{l_\down} = e^{-2\eta\alpha^2}\ .
\end{equation}
We reflect the light when the atom is in an equal superposition of $\ket{\up}$ and $\ket{\down}$ and probe the atom after a consecutive $\pi/2$ rotation with phase $\theta$. The produced optical state immediately after the reflection and state detection of the atom is
\begin{equation}
\ket{\psi_\textnormal{out}} = \frac{\ket{r_\up}\ket{l_\up}+e^{i\theta}\ket{r_\down}\ket{l_\down}}{\sqrt{2(1+e^{-2\eta\alpha^2}\cos\theta)}}
\end{equation}
for the atom measured in $\ket{\down}$, and similar with $\theta\rightarrow\theta+\pi$ for the atom in $\ket{\up}$. Light in the loss modes will be dissipated in the environment. The remaining optical state $\rho$ is obtained by tracing out the losses:
\begin{equation}
\rho = \operatorname{tr}_{l} \ket{\psi_\textnormal{out}}\bra{\psi_\textnormal{out}} 
\end{equation}
\begin{eqnarray}
= &\frac{1/2}{1+e^{-2\eta\alpha^2}\cos\theta}&\Bigl(
\ket{r_\up}\bra{r_\up}\normt{l_\up}{l_\up}
+ e^{-i\theta}\ket{r_\up}\bra{r_\down}\normt{l_\down}{l_\up}+\nonumber\\
&&
{}+ e^{i\theta}\ket{r_\down}\bra{r_\up}\normt{l_\up}{l_\down}
+\ket{r_\down}\bra{r_\down}\normt{l_\down}{l_\down}
\Bigr)\qquad
\end{eqnarray}
\begin{equation}
= \frac{
  \ket{r_\up}\bra{r_\up}
+ e^{-2(1-\eta)\eta\alpha^2}(e^{-i\theta}\ket{r_\up}\bra{r_\down}
+ e^{i\theta}\ket{r_\down}\bra{r_\up})
+ \ket{r_\down}\bra{r_\down}
}{2(1+e^{-2\eta\alpha^2}\cos\theta)}
\end{equation}
\begin{equation}
= \frac{
  \ket{r_\up}\bra{r_\up}
+ e^{-2(1-\eta)\alpha_0^2}(e^{-i\theta}\ket{r_\up}\bra{r_\down}
+ e^{i\theta}\ket{r_\down}\bra{r_\up})
+ \ket{r_\down}\bra{r_\down}
}{2(1+e^{-2\alpha_0^2}\cos\theta)}
\end{equation}
Here, the coherence terms are reduced due to optical losses by a factor $\exp(-2(1-\eta)\alpha_0^2)$. The state $\rho$ is equivalent to a cat state of original amplitude $\alpha_0=\sqrt{\eta}\alpha$ that has undergone coherence-reducing intensity losses \cite{spagnolo2009} of $L_\textnormal{eff}=1-\eta\ (=0.19)$. Thus, only a part of the total cavity losses $L_\textnormal{cav}=1-\eta^2$ affects the coherences. In terms of optical depth, the coherence-reducing losses are exactly half the total losses.

The Wigner function $W$ of $\rho$, given for a generic lossy cat state in ref.~\cite{spagnolo2009}, is
\begin{eqnarray}
W(q,p) &=& \frac{1}{2\pi}\Bigl(
e^{-p^2-(q-\sqrt{2}r_{\down})^2}
{}+e^{-p^2-(q-\sqrt{2}r_{\up})^2}\nonumber\\
&&+2\,e^{-2(1-\eta)\eta\alpha^2}\,e^{-p^2-(q-(r_{\down}+r_{\up})/\sqrt{2})^2}\nonumber\\
&&\cdot\cos(\theta + \sqrt{8}\eta\alpha p)\Bigr)/(1+e^{-2\eta\alpha^2}\cos\theta)
\end{eqnarray}
It consists of two Gaussian peaks of separation $\sqrt{8}\eta\alpha$ and a fringe term in the centre whose amplitude is reduced by $\exp(-2(1-\eta)\alpha_0^2)$ through the losses. The fringe centre $(q_0,p_0)=((r_{\down}+r_{\up})/\sqrt{2}, 0)$, is equal to $(0,0)$ in our experiment, because the reflection amplitudes $r_{\down}$ and $r_{\up}$ have the same magnitude. The interference fringe visibility defined by $V=\frac\pi2(W_\textnormal{even}(q_0, p_0)-W_\textnormal{odd}(q_0, p_0))$ becomes
\begin{equation}
V = \frac{\sinh(2\eta\alpha_0^2)}{\sinh(2\alpha_0^2)}= \frac{\sinh(2\left(1-L_{\textnormal{eff}}\right)\alpha_0^2)}{\sinh(2\alpha_0^2)},
\end{equation}
taking into account the normalization of $W$. If the coherent contributions in the final cat states have little overlap, $\eta\alpha_0^2 \gg 1$, the visibility decays exponentially with respect to the losses and to the cat size $\alpha_0^2$:
\begin{equation}
V \simeq \exp(-2L_\textnormal{eff}\alpha_0^2).
\end{equation}
The fringes essentially vanish when the number of lost photons $L_\textnormal{eff}\alpha_0^2$ exceeds one half. This effectively limits the achievable size of a cat state.

Finally we compute the fidelity of $\rho$ with respect to an ideal cat state of the same size
\begin{equation}
\ket{\psi_\textnormal{cat}} =
\frac{\ket{r_\up}+e^{i\theta}\ket{r_\down}}{\sqrt{2(1+e^{-2\eta^2\alpha^2}\cos\theta)}}
\end{equation}
which yields
\begin{equation}
F = \bra{\psi_\textnormal{cat}}\rho\ket{\psi_\textnormal{cat}}
\end{equation}
\begin{equation}
= 1 - \frac{(1-e^{-4\eta\alpha_0^2})(1-e^{-2(1-\eta)\alpha_0^2})}
{2(1+e^{-2\eta\alpha_0^2}\cos\theta)(1+e^{-2\alpha_0^2}\cos\theta)}.
\end{equation}

\clearpage

\setcounter{figure}{0}

\renewcommand{\figurename}{\textbf{Fig.}}
\renewcommand{\thefigure}{\arabic{figure}}
\renewcommand{\tablename}{\textbf{Table}}
\renewcommand{\thetable}{\arabic{table}}

\makeatletter
\renewcommand{\fnum@figure}{\figurename~\textbf{\thefigure}}
\renewcommand{\fnum@table}{\tablename~\textbf{\thetable}}
\makeatother

\onecolumngrid
{\centering
\textbf{\large Supplementary Information: Deterministic creation of entangled atom-light Schr{\"o}dinger-cat states}\\[\baselineskip]
Bastian~Hacker, Stephan~Welte, Severin~Daiss, Armin~Shaukat, Stephan~Ritter, Lin~Li, Gerhard~Rempe\\
\textit{\small Max-Planck-Institut f\"ur Quantenoptik, Hans-Kopfermann-Strasse 1, 85748 Garching, Germany}\\[2.5\baselineskip]}
\twocolumngrid

\renewcommand\thesection{S\arabic{section}}
\setcounter{section}{0}
\renewcommand\theequation{S\arabic{equation}}
\setcounter{equation}{0}
\renewcommand\thefigure{S\arabic{figure}}
\setcounter{figure}{0}
\renewcommand\thetable{S\arabic{table}}
\setcounter{table}{0}

\section{Loss budget}
\label{supplement:loss_budget}
\begin{table}[b]
\caption{\label{tab:losses}
Loss budget listing the loss channels in our experiment that we individually quantified. Uncertainties in the individual losses are on the order of 10\% of their respective magnitude.}
\begin{ruledtabular}%
\begin{tabular}{lrl@{}l}%
Source of loss & \hspace{-3em}(effective) loss & $L_i$ & \\
\hline
Finite cavity reflectivity & 19.0\% & $L_\textnormal{eff}$ & \\
Lenses, waveplates, mirrors and NPBS & 9.5\% & \rdelim\}{3.5}{*}[$L_\textnormal{o}$] &
\rdelim\}{10.8}{*}[$L_\textnormal{p+d}$]\\
Isolator transmission & 3.0\% & & \\
Switch AOD transmission & 2.5\% & & \\
Mode matching with LO & 6.0\% & $L_\textnormal{m}$ & \\
Quantum efficiency of photodiodes & 1.5\% & $L_\textnormal{d}$ & \\
Detector dark noise & 2.5\% & \rdelim\}{4.6}{*}[$L_\textnormal{n}$] & \\
LO classical laser noise & 1.8\% & & \\
Electronic high pass 0.7\,kHz signal reduction & 1.1\% & & \\
Electronic high pass background noise & 0.2\% & & \\
\hline
Total losses & 39.3\% & &
\end{tabular}%
\end{ruledtabular}%
\end{table}
A detailed list of all identified loss sources in our set-up is shown in Table~\ref{tab:losses}.
Cavity losses $L_\textnormal{eff}=19\%$ were inferred from the measured cavity-QED parameters, as shown in the methods. They are the only intrinsic losses, whereas everything else can be considered technical.

Optical absorption and reflection losses amount to $L_\textnormal{o}=14\%$ and are induced by the optical isolator and the acousto-optical deflector (AOD) switch as well as other optical components such as lenses, waveplates, mirrors and beamsplitters. Those losses were determined with a power meter at a macroscopic beam intensity. The imperfect mode matching between the signal and the local oscillator was determined through the visibility of an interference signal. The quantum efficiency of the two homodyne photodiodes was determined from the photocurrent at a given optical power. The various noise sources along the signal line, optical as well as electronic, resemble optical losses $L_\textnormal{n}$ that can be derived from the noise variances \cite{appel2007_supp}.

When several losses $L_i$ occur, their transmission coefficients multiply, that is the combined loss $L$ is obtained according to
\begin{equation}
1-L=\prod_i(1-L_{i})\;.
\end{equation}
We obtain combined propagation and detection losses $L_\textnormal{p+d}=25\%$, which may be corrected for in the evaluation. Together with $L_\textnormal{eff}$, all losses that we individually accounted for (Table~\ref{tab:losses}) amount to $39\%$. The experimentally measured fringe visibilities correspond to slightly higher actual losses of $L_\textnormal{det}=46\%$ (Fig.~3b). This is mostly caused by two imperfections: Fluctuations of the cavity-induced phase shift ($\Delta \phi=0.06\pi$) blur the Wigner function and reduce the amplitude of the interference fringes. This effect reduces the visibility as much as another $5\%$ loss for a cat state with $\alpha=1.4$. In addition, imperfect atomic state detection causes a mutual admixture of opposite cat states that partially cancels the interference fringes. We determined the fraction of wrong atomic state detections, including imperfect spin rotations, when no light was impinging to $1.3\%$. This reduces the fringe visibility at $\alpha=1.4$ as much as another $1\%$ loss.

We expect that the technical losses could be reduced significantly by a multitude of measures. For instance, lenses could be replaced by curved mirrors and the beam-switching could be performed by a more efficient device than the acousto-optical deflector. All optical elements could be selected more rigorously for low optical losses, as the standard off-the-shelf components exhibit significant performance variations.

\section{Squeezing of cat states}
\label{supplement:squeezing}
\begin{figure}[tb]
\centering
\hypertarget{fig:squeezing-wigner-histo}{}
\includegraphics[width=8.6cm]{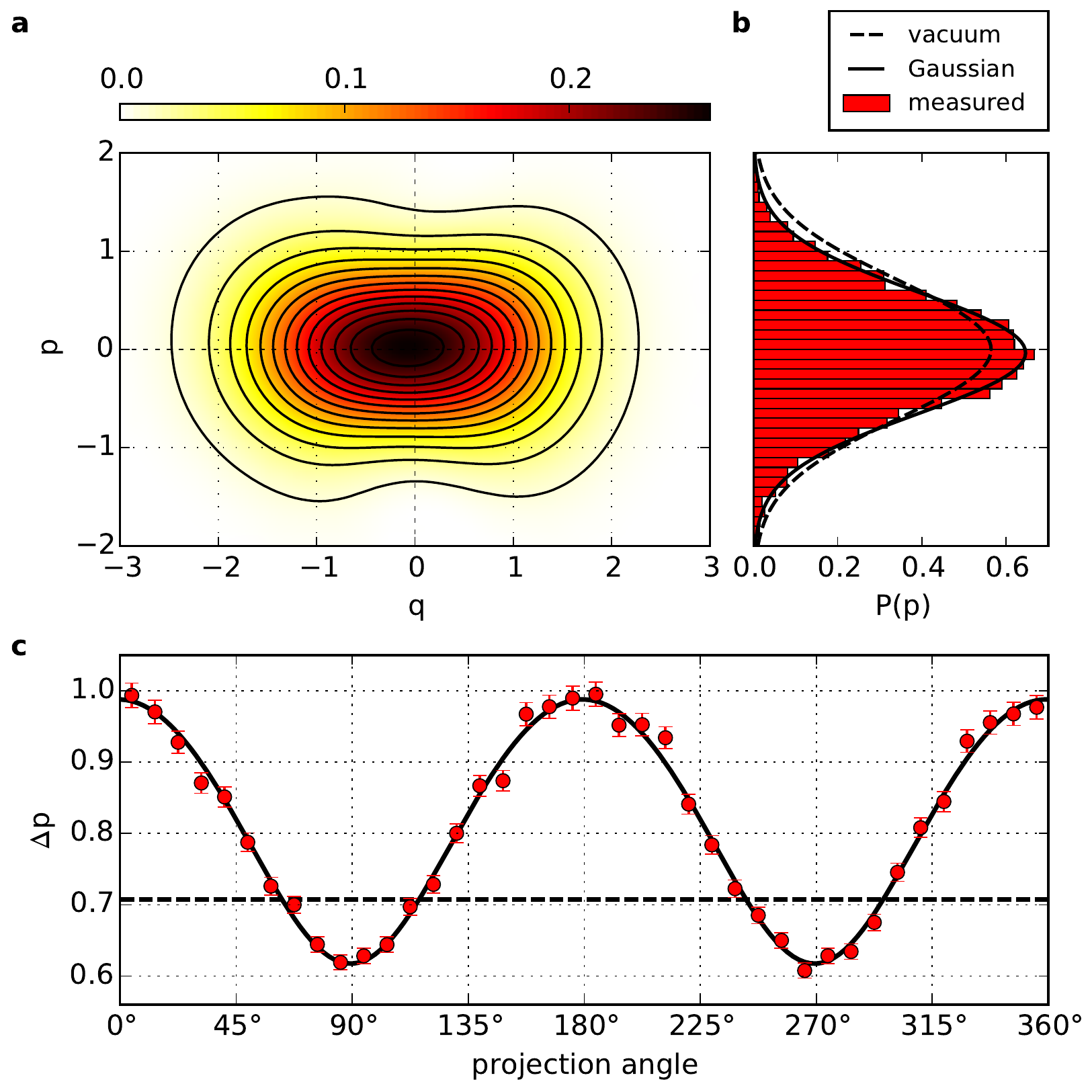}
\caption{\label{fig:squeezing-wigner-histo}
\textbf{Squeezed cat state at $\boldsymbol{\alpha=0.7}$.}
\textbf{a}, Reconstructed Wigner distribution from $\approx7{\times}10^4$ measured quadrature values by means of the maximum likelihood estimation in Fock space. No loss-correction was applied.
\textbf{b}, Histogram of measured quadrature values for projection angles in the interval $90^\circ{\pm}5^\circ$. The solid curve is a normalized Gaussian with the same mean value and spread as the measured distribution. The dashed curve shows the respective distribution of a vacuum state.
\textbf{c}, Standard deviation of quadrature values vs.\ projection angle given by the relative phase of the local oscillator. Fitted values range from $0.62$ (squeezed) to $0.99$ (anti-squeezed). The vacuum noise level is at $1/\sqrt{2}$ (dashed line).}
\end{figure}

Fig.~\figref{fig:squeezing-wigner-histo} shows an example of an even cat state of $\alpha=0.7$ and the respective histogram along the $p$-axis. The data show squeezing of $1.18(3)\,\mathrm{dB}$.
Squeezing can only be observed for even cat states. The odd cat states have a minimum at the centre of phase space which necessarily broadens the distribution. Fig.~3a in the main text shows a scan of the cat size and the corresponding width of the respective distributions for the even and the odd cat states.

\section{Control over all parameters of the cat states}
\label{supplement:control-cat}
\begin{figure*}[p]
\centering
\setlength{\unitlength}{\textwidth}
\begin{picture}(1,0.94)
\put(0,0.62){\includegraphics[width=\textwidth]{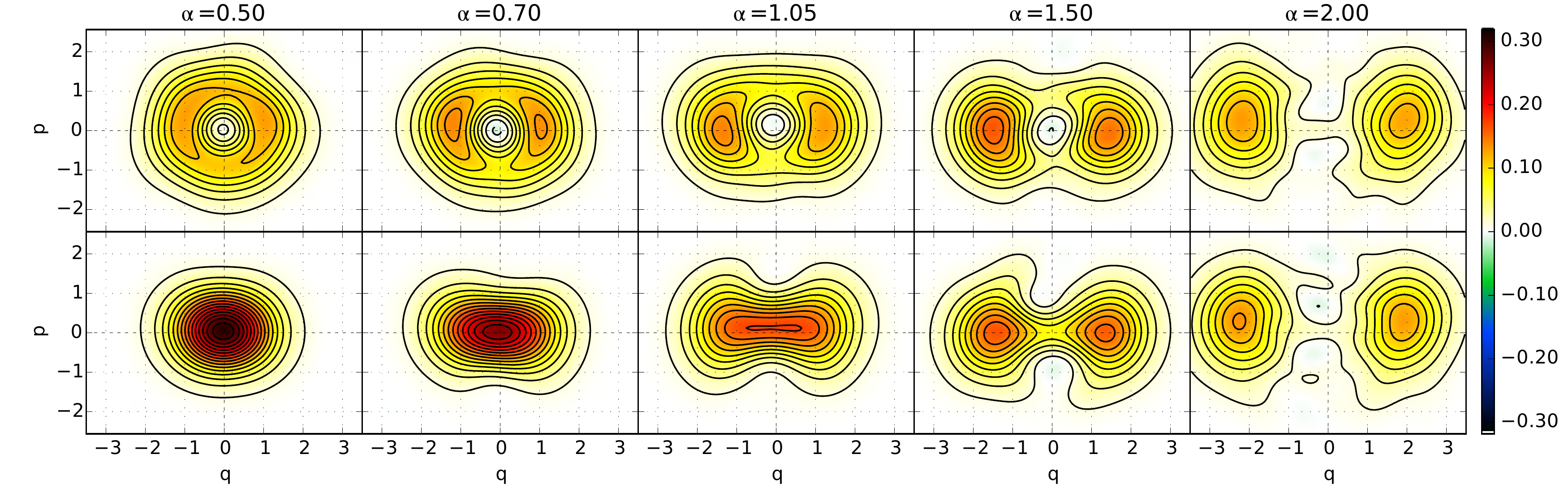}}
\put(0,0.32){\includegraphics[width=\textwidth]{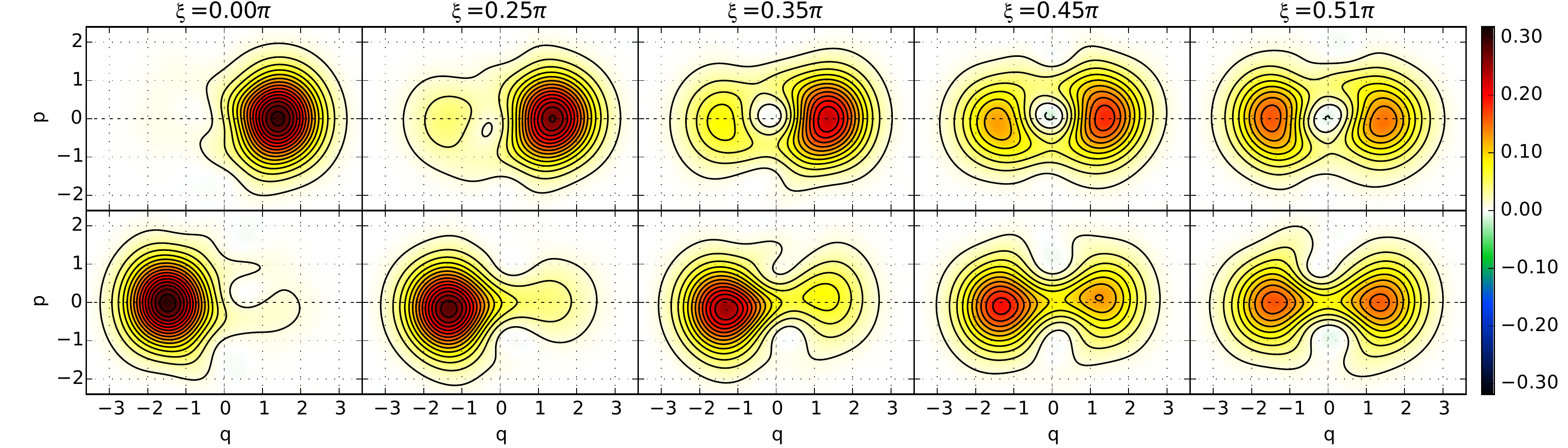}}
\put(0,0){\includegraphics[width=\textwidth]{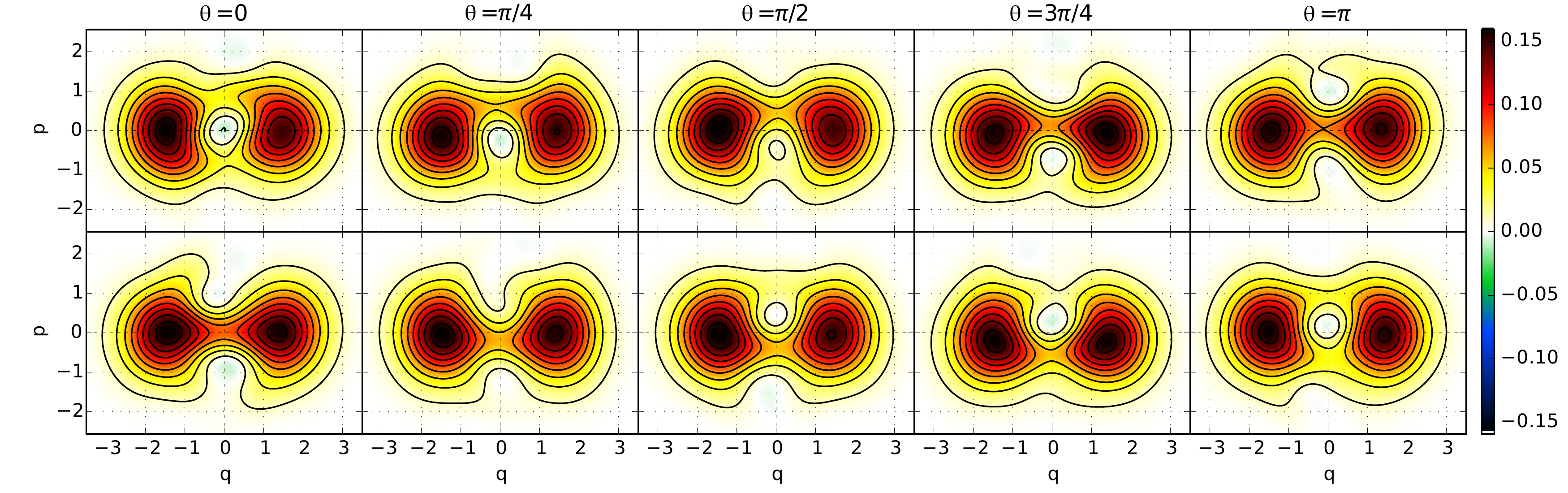}}
\put(0,0.92){\textbf{a}\hypertarget{fig:scanalpha_supp}{}}
\put(0,0.60){\textbf{b}\hypertarget{fig:scanxi_supp}{}}
\put(0,0.31){\textbf{c}\hypertarget{fig:scantheta_supp}{}}
\end{picture}
\caption{\label{fig:scanalpha_supp}\label{fig:scanxi_supp}\label{fig:scantheta_supp}
\textbf{Gallery of cat states.} Each top row displays the optical Wigner functions for the atom detected in $\ket{\up}$ whereas the bottom row is for $\ket{\down}$.
\textbf{a}, Odd and even cat states for varying amplitudes $\alpha$ between $0.5$ and $2$. The most significant negative values are observed for $\alpha=1.4$. Above $\alpha=2$ the interference fringes become dominated by noise.
\textbf{b}, Wigner functions for varying pulse area $\xi$ of the final spin rotation pulse (step 4 in Fig.~2a). We can smoothly transition between coherent states ($\xi=0$) and cat states ($\xi=\pi/2$). Here $\alpha=1.4$.
\textbf{c}, Scan of the phase $\theta$. A continuous transition from an odd into an even cat state and vice versa is observable. The generated cat states have the form $(\ket{\alpha}\mp e^{i\theta}\ket{-\alpha})/\mathcal{N}$ with $\alpha=1.4$.
}
\end{figure*}
In this section, we demonstrate full control over all parameters that quantify the cat state
\begin{equation}
\ket{\psi_{\textnormal{cat}}} = \frac{1}{\mathcal{N}} \left(\cos(\xi/2)\ket{\alpha} + e^{i\theta}\sin(\xi/2)\ket{e^{i\phi}\alpha}\right)
.
\end{equation}
Specifically, we scan the size $\alpha$, the optical phase $\phi$ between the coherent contributions, the superposition phase $\theta$ and population fraction of the two coherent contributions $\xi$.

\subsection{Scan of cat-state amplitude \texorpdfstring{$\boldsymbol{\alpha}$}{$\alpha$}}
\label{supplement:cat-size}
The cat state amplitude $\alpha$ can be chosen freely by the intensity of the coherent input light which is proportional to the mean number of photons $\langle{n}\rangle=\vert\alpha\vert^2$. Fig.~\figref{fig:scanalpha_supp}a shows odd and even cat states for different $\alpha$.
If $\alpha$ is increased in the experiment, the two coherent Gaussian peaks in the Wigner function separate further. Additionally, the oscillation period of the observed interference fringes between these contributions scales proportional to $\alpha$. The even cat states show quadrature squeezing for $\alpha<1.1$. Due to optical losses, the interference fringes and the negativity drop to the noise level if the amplitude is increased above $\alpha=2$.

\begin{figure*}[tb]
\centering
\hypertarget{fig:phaseshift-wigner}{}
\includegraphics[width=\textwidth]{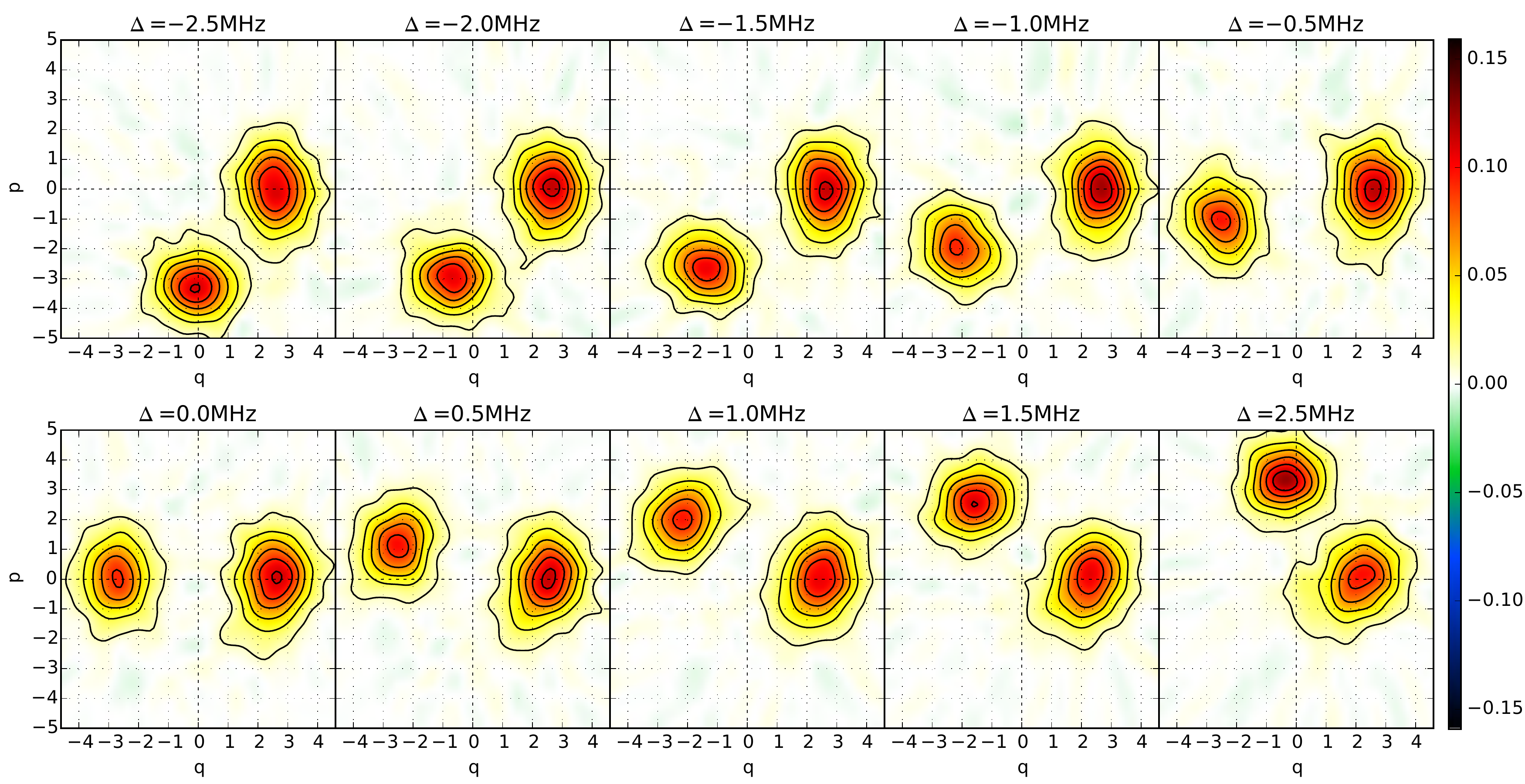}
\caption{\label{fig:phaseshift-wigner}
\textbf{Reflected coherent pulses with cavity-light detuning $\boldsymbol{\Delta}$.} Wigner functions of reflected coherent pulses of $\alpha=2.3$ with the atom prepared in $(\ket{\up}+\ket{\down})/\sqrt{2}$ and no consideration of the final atomic state. The detuning $\Delta$ between the impinging light and the cavity is scanned between $\pm2.5\,\mathrm{MHz}$ in steps of $0.5\,\mathrm{MHz}$. The Gaussian on the positive $q$ axis corresponds to the coupling atom $\ket{\up}$ while the other Gaussian corresponding to $\ket{\down}$ moves in phase space as $\Delta$ is changed. Because $\alpha$ is relatively large and the final atomic state is not considered, no interference fringes appear. The Wigner functions in this figure were reconstructed with the inverse Radon transform, because it is better suited for states with large amplitudes than the maximum likelihood technique.}
\end{figure*}
\begin{figure}[tb]
\centering
\hypertarget{fig:phaseshift-angle}{}
\includegraphics[width=\columnwidth]{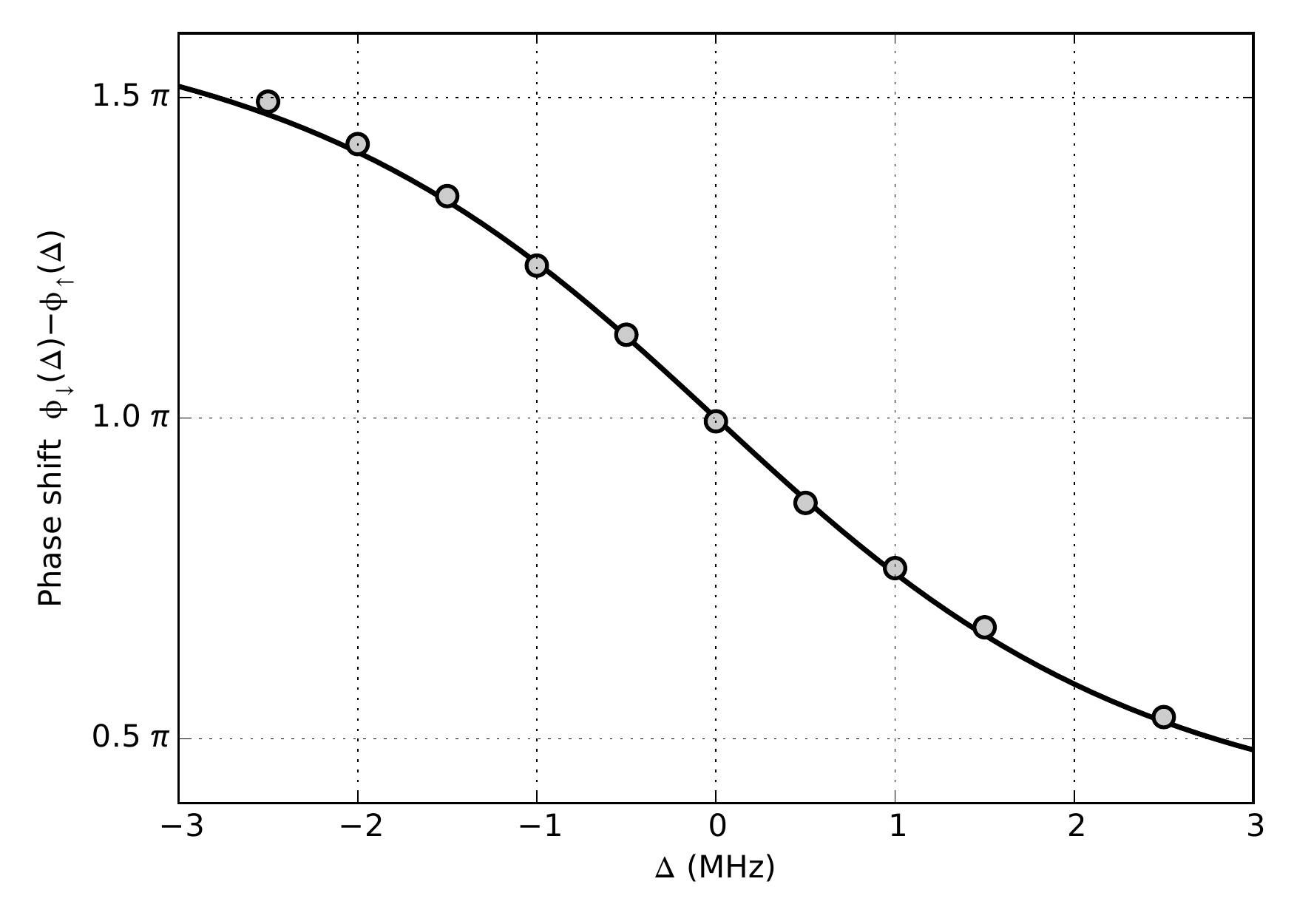}
\caption{\label{fig:phaseshift-angle}
\textbf{Phaseshift $\boldsymbol{\phi}$ as a function of cavity-light detuning $\boldsymbol{\Delta}$.} Atom-induced phase shift as a function of laser detuning. The black solid curve shows $\phi(\Delta)$ according to equation~(\ref{eq:phiofdelta}) with our cavity parameters. The measured angles, obtained from fits to the Wigner functions in Fig.~\protect\figref{fig:phaseshift-wigner}, closely follow the theoretical prediction. Error bars on each data point are in the range of one degree and thus too small to be visible in the above plot.}
\end{figure}

\subsection{Scan of coherent population fraction \texorpdfstring{$\boldsymbol{\xi}$}{$\xi$}}
\label{supplement:scan-population}
The ratio of the two coherent contributions $\ket{\alpha}$ and $\ket{-\alpha}$ in the produced optical states can be controlled by changing the area $\xi$ of the final spin rotation of the atom (step 4 in Fig.~2a). Fig.~\figref{fig:scanxi_supp}b shows such a scan for different values of $\xi$ between $0$ and $\pi/2$. The experiment performs a continuous transition between the entangled Schr{\"o}dinger-cat states $\bigl(\ket{\up}\ket{\alpha}+\ket{\down}\ket{-\alpha}\bigr)/\sqrt{2}$ and $\bigl[\ket{\up}\bigl(\ket{\alpha}-\ket{-\alpha}\bigr)+\ket{\down}\bigl(\ket{\alpha}+\ket{-\alpha}\bigr)\bigr]/2$.

\subsection{Scan of the superposition phase \texorpdfstring{$\boldsymbol{\theta}$}{$\theta$}}
\label{supplement:scan-fringes}
The interference fringes in the Wigner function can be controlled via a change of the phase $\theta$ of the last $\pi/2$ spin rotation in the experimental protocol. In all previous experiments, the phase of this particular pulse was equal to the phase of the first $\pi/2$ pulse. Experimentally, the phase of the last pulse can be altered by a change in the phase of the respective radio frequency drive supplying an acousto-optical modulator which generates the desired Raman probe pulse. We employ a direct digital synthesizer to scan this phase from zero to $\pi$ in steps of $\pi/4$. Fig.~\figref{fig:scantheta_supp}c shows the maximum likelihood reconstructions of the respective Wigner functions. For $\theta=0$, a post-selection on $\ket{\uparrow}$ results again in the odd cat state. In the case of the atom being measured in $\ket{\down}$, the even cat state is obtained. This correspondence can be interchanged by setting $\theta=\pi$ as done in the last column of Fig.~\figref{fig:scantheta_supp}c. In the intermediate regime \cite{yurke1986_supp}, a continuous transition from the even to the odd cat state and vice versa can be observed when post-selecting on $\ket{\down}$ or $\ket{\up}$, respectively. We observe a direct correspondence of the measured fringe phase $\theta$ to the respective phase of the atomic state rotation pulse.

\subsection{Scan of the optical phase \texorpdfstring{$\boldsymbol{\phi}$}{$\phi$}}
\label{supplement:phaseshift}
To characterize the phase shift mechanism employed to create the cat states, the atom is prepared in the superposition state $(\ket{\up}+\ket{\down})/\sqrt{2}$. Subsequently, a resonant coherent pulse with $\alpha=2.3$ ($\langle{n}\rangle=5.3$) is reflected from the cavity. The reflected pulse is tomographically characterized and the respective Wigner function is reconstructed. The Wigner function shows the two Gaussian distributions whose relative positions depend on the imprinted phase shift and thus on the relative detuning $\Delta$ of the coherent pulse and the cavity resonance \cite{duan2004_supp, kalb2015_supp}. Cavity input-output theory predicts the respective complex reflection amplitude $r$ as
\begin{equation}
\label{eq:reflectivity}
r(\Delta)=1-\frac{2\kappa_r(2i\pi\Delta+\gamma)}{(2i\pi\Delta+\kappa)(2i\pi\Delta+\gamma)+g^2}.
\end{equation}  
The relative phase shift between the coupling and the non-coupling case is thus given by 
\begin{equation}
\label{eq:phiofdelta}
\phi(\Delta)=\arg(r(\Delta)\vert_{g=0})-\arg(r(\Delta)\vert_{g=2\pi\times7.8\text{MHz}}).
\end{equation}
In the phase space representation, the relative phase shift can be extracted as the angle between the lines connecting the origin to the two centres of the observed Gaussians. Thus, the angular distance between the two Gaussians will vary as $\Delta$ and thus $\phi(\Delta)$ is scanned (Fig.~\figref{fig:phaseshift-wigner}). The amplitude of the coherent states varies slightly for different $\Delta$ as the modulus of the reflectivity (equation~(\ref{eq:reflectivity})) changes across the cavity resonance. Fig.~\figref{fig:phaseshift-angle} shows the theoretical expectation for $\phi(\Delta)$ with our cavity parameters as well as the experimentally observed data. We find agreement between theory and experiment.

\section{Detuned cat states}
\label{supplement:detuned-cat}
\begin{figure}[b]
\centering
\hypertarget{fig:detuned-cat}{}
\includegraphics[width=\columnwidth]{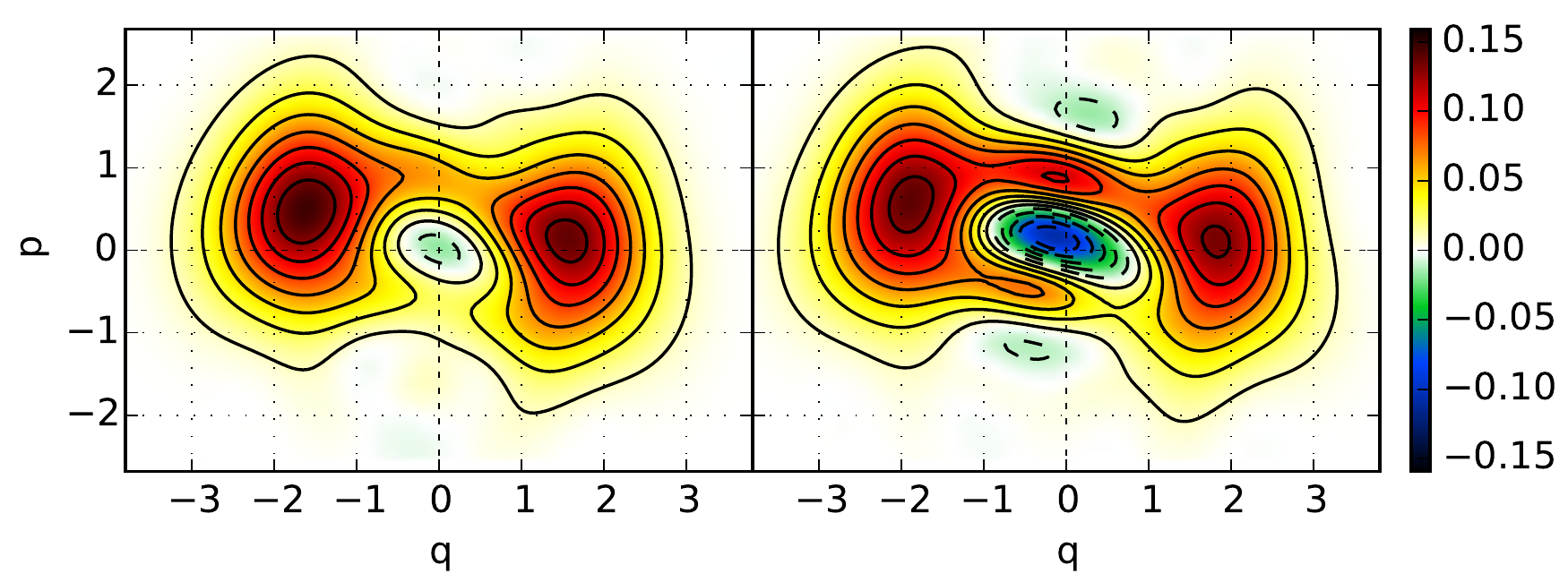}
\caption{\label{fig:detuned-cat}
\textbf{Example of a cat state generated with a detuned coherent pulse reflected from the atom-cavity system.}
The left plot shows the Wigner function reconstructed from the raw data while the right plot shows the same Wigner function corrected for propagation and detection losses. The detuning between the cavity resonance and the optical pulse is $\Delta=300\,\mathrm{kHz}$.}
\end{figure}
Experimentally, $\phi$ can be continuously tuned between $0$ and $2\pi$ by changing the resonance frequency of the cavity. This degree of freedom can be used to generate cat states in which the two contributions have any desired phase difference in phase space. Fig.~\figref{fig:detuned-cat} shows an example of an odd cat state where a detuning of $\Delta=300\,\mathrm{kHz}$ was chosen. In this case, the phase shift between the non-coupling and the coupling component of the reflected light is $0.91\pi$.

The ability to control the phase difference offers the possibility to create cat states with more than two coherent contributions after repeated reflections. If a detuned cat state is reflected from another cavity and if the detuning is changed between the two reflection processes, (this could be done by locking the cavities at different frequencies) a cat state with four different coherent components can be created. Such cat states have been generated in the microwave domain \cite{heeres2017_supp} and were shown to be a useful resource in error correction schemes \cite{leghtas2013_supp}.


\begin{thebibliography}{10}
\expandafter\ifx\csname url\endcsname\relax
  \def\url#1{\texttt{#1}}\fi
\expandafter\ifx\csname urlprefix\endcsname\relax\def\urlprefix{URL }\fi
\providecommand{\bibinfo}[2]{#2}
\providecommand{\eprint}[2][]{\url{#2}}

\bibitem{schrodinger1935}
\bibinfo{author}{Schr{\"o}dinger, E.}
\newblock \bibinfo{title}{Die gegenw{\"a}rtige Situation in der
  Quantenmechanik}.
\newblock
  \href{https://www.doi.org/10.1007/BF01491891}{\textit{\bibinfo{journal}{Naturwissenschaften}}
  \textbf{\bibinfo{volume}{23}}\textbf{,} \bibinfo{pages}{807--812}
  (\bibinfo{year}{1935})}.

\bibitem{glancy2008}
\bibinfo{author}{Glancy, S.} \& \bibinfo{author}{Vasconcelos, H. M.~d.}
\newblock \bibinfo{title}{Methods for producing optical coherent state
  superpositions}.
\newblock
  \href{https://www.doi.org/10.1364/JOSAB.25.000712}{\textit{\bibinfo{journal}{J.
  Opt. Soc. Am. B}} \textbf{\bibinfo{volume}{25}}\textbf{,}
  \bibinfo{pages}{712--733} (\bibinfo{year}{2008})}.

\bibitem{wineland2013}
\bibinfo{author}{Wineland, D.~J.}
\newblock \bibinfo{title}{Nobel Lecture: Superposition, entanglement, and
  raising Schr{\"o}dinger's cat}.
\newblock
  \href{https://www.doi.org/10.1103/RevModPhys.85.1103}{\textit{\bibinfo{journal}{Rev.
  Mod. Phys.}} \textbf{\bibinfo{volume}{85}}\textbf{,}
  \bibinfo{pages}{1103--1114} (\bibinfo{year}{2013})}.

\bibitem{kienzler2016}
\bibinfo{author}{Kienzler, D.} \emph{et~al.}
\newblock \bibinfo{title}{Observation of Quantum Interference between Separated
  Mechanical Oscillator Wave Packets}.
\newblock
  \href{https://www.doi.org/10.1103/PhysRevLett.116.140402}{\textit{\bibinfo{journal}{Phys.
  Rev. Lett.}} \textbf{\bibinfo{volume}{116}}\textbf{,} \bibinfo{pages}{140402}
  (\bibinfo{year}{2016})}.

\bibitem{deleglise2008}
\bibinfo{author}{Deleglise, S.} \emph{et~al.}
\newblock \bibinfo{title}{Reconstruction of non-classical cavity field states
  with snapshots of their decoherence}.
\newblock
  \href{https://www.doi.org/10.1038/nature07288}{\textit{\bibinfo{journal}{Nature}}
  \textbf{\bibinfo{volume}{455}}\textbf{,} \bibinfo{pages}{510--514}
  (\bibinfo{year}{2008})}.

\bibitem{haroche2013}
\bibinfo{author}{Haroche, S.}
\newblock \bibinfo{title}{Nobel Lecture: Controlling photons in a box and
  exploring the quantum to classical boundary}.
\newblock
  \href{https://www.doi.org/10.1103/RevModPhys.85.1083}{\textit{\bibinfo{journal}{Rev.
  Mod. Phys.}} \textbf{\bibinfo{volume}{85}}\textbf{,}
  \bibinfo{pages}{1083--1102} (\bibinfo{year}{2013})}.

\bibitem{vlastakis2013}
\bibinfo{author}{Vlastakis, B.} \emph{et~al.}
\newblock \bibinfo{title}{Deterministically {E}ncoding {Q}uantum {I}nformation
  {U}sing 100-{P}hoton Schr{\"o}dinger {C}at {S}tates}.
\newblock
  \href{https://www.doi.org/10.1126/science.1243289}{\textit{\bibinfo{journal}{Science}}
  \textbf{\bibinfo{volume}{342}}\textbf{,} \bibinfo{pages}{607--610}
  (\bibinfo{year}{2013})}.

\bibitem{pfaff2017}
\bibinfo{author}{Pfaff, W.} \emph{et~al.}
\newblock \bibinfo{title}{Controlled release of multiphoton quantum states from
  a microwave cavity memory}.
\newblock
  \href{https://www.doi.org/10.1038/nphys4143}{\textit{\bibinfo{journal}{Nat.
  Phys.}} \textbf{\bibinfo{volume}{13}}\textbf{,} \bibinfo{pages}{882--887}
  (\bibinfo{year}{2017})}.

\bibitem{morin2014}
\bibinfo{author}{Morin, O.} \emph{et~al.}
\newblock \bibinfo{title}{Remote creation of hybrid entanglement between
  particle-like and wave-like optical qubits}.
\newblock
  \href{https://www.doi.org/10.1038/nphoton.2014.137}{\textit{\bibinfo{journal}{Nat.
  Photon.}} \textbf{\bibinfo{volume}{8}}\textbf{,} \bibinfo{pages}{570--574}
  (\bibinfo{year}{2014})}.

\bibitem{jeong2014}
\bibinfo{author}{Jeong, H.} \emph{et~al.}
\newblock \bibinfo{title}{Generation of hybrid entanglement of light}.
\newblock
  \href{https://www.doi.org/10.1038/nphoton.2014.136}{\textit{\bibinfo{journal}{Nat.
  Photon.}} \textbf{\bibinfo{volume}{8}}\textbf{,} \bibinfo{pages}{564--569}
  (\bibinfo{year}{2014})}.

\bibitem{ulanov2017}
\bibinfo{author}{Ulanov, A.~E.}, \bibinfo{author}{Sychev, D.},
  \bibinfo{author}{Pushkina, A.~A.}, \bibinfo{author}{Fedorov, I.~A.} \&
  \bibinfo{author}{Lvovsky, A.~I.}
\newblock \bibinfo{title}{Quantum Teleportation Between Discrete and Continuous
  Encodings of an Optical Qubit}.
\newblock
  \href{https://www.doi.org/10.1103/PhysRevLett.118.160501}{\textit{\bibinfo{journal}{Phys.
  Rev. Lett.}} \textbf{\bibinfo{volume}{118}}\textbf{,} \bibinfo{pages}{160501}
  (\bibinfo{year}{2017})}.

\bibitem{jeannic2018}
\bibinfo{author}{Jeannic, H.~L.}, \bibinfo{author}{Cavaill\`{e}s, A.},
  \bibinfo{author}{Raskop, J.}, \bibinfo{author}{Huang, K.} \&
  \bibinfo{author}{Laurat, J.}
\newblock \bibinfo{title}{Remote preparation of continuous-variable qubits
  using loss-tolerant hybrid entanglement of light}.
\newblock
  \href{https://www.doi.org/10.1364/OPTICA.5.001012}{\textit{\bibinfo{journal}{Optica}}
  \textbf{\bibinfo{volume}{5}}\textbf{,} \bibinfo{pages}{1012--1015}
  (\bibinfo{year}{2018})}.

\bibitem{ourjoumtsev2006}
\bibinfo{author}{Ourjoumtsev, A.}, \bibinfo{author}{Tualle-Brouri, R.},
  \bibinfo{author}{Laurat, J.} \& \bibinfo{author}{Grangier, P.}
\newblock \bibinfo{title}{Generating Optical Schr{\"o}dinger Kittens for
  Quantum Information Processing}.
\newblock
  \href{https://www.doi.org/10.1126/science.1122858}{\textit{\bibinfo{journal}{Science}}
  \textbf{\bibinfo{volume}{312}}\textbf{,} \bibinfo{pages}{83--86}
  (\bibinfo{year}{2006})}.

\bibitem{ourjoumtsev2007}
\bibinfo{author}{Ourjoumtsev, A.}, \bibinfo{author}{Jeong, H.},
  \bibinfo{author}{Tualle-Brouri, R.} \& \bibinfo{author}{Grangier, P.}
\newblock \bibinfo{title}{Generation of optical
  {\textquoteleft}{Schr{\"o}dinger} cats{\textquoteright} from photon number
  states}.
\newblock
  \href{https://www.doi.org/10.1038/nature06054}{\textit{\bibinfo{journal}{Nature}}
  \textbf{\bibinfo{volume}{448}}\textbf{,} \bibinfo{pages}{784}
  (\bibinfo{year}{2007})}.

\bibitem{nielsen2006}
\bibinfo{author}{Neergaard-Nielsen, J.~S.}, \bibinfo{author}{Nielsen, B.~M.},
  \bibinfo{author}{Hettich, C.}, \bibinfo{author}{M{\o}lmer, K.} \&
  \bibinfo{author}{Polzik, E.~S.}
\newblock \bibinfo{title}{Generation of a Superposition of Odd Photon Number
  States for Quantum Information Networks}.
\newblock
  \href{https://www.doi.org/10.1103/PhysRevLett.97.083604}{\textit{\bibinfo{journal}{Phys.
  Rev. Lett.}} \textbf{\bibinfo{volume}{97}}\textbf{,} \bibinfo{pages}{083604}
  (\bibinfo{year}{2006})}.

\bibitem{takahashi2008}
\bibinfo{author}{Takahashi, H.} \emph{et~al.}
\newblock \bibinfo{title}{Generation of Large-Amplitude Coherent-State
  Superposition via Ancilla-Assisted Photon Subtraction}.
\newblock
  \href{https://www.doi.org/10.1103/PhysRevLett.101.233605}{\textit{\bibinfo{journal}{Phys.
  Rev. Lett.}} \textbf{\bibinfo{volume}{101}}\textbf{,} \bibinfo{pages}{233605}
  (\bibinfo{year}{2008})}.

\bibitem{lvovsky2009}
\bibinfo{author}{Lvovsky, A.~I.} \& \bibinfo{author}{Raymer, M.~G.}
\newblock \bibinfo{title}{Continuous-variable optical quantum-state
  tomography}.
\newblock
  \href{https://www.doi.org/10.1103/RevModPhys.81.299}{\textit{\bibinfo{journal}{Rev.
  Mod. Phys.}} \textbf{\bibinfo{volume}{81}}\textbf{,}
  \bibinfo{pages}{299--332} (\bibinfo{year}{2009})}.

\bibitem{namekata2010}
\bibinfo{author}{Namekata, N.} \emph{et~al.}
\newblock \bibinfo{title}{Non-Gaussian operation based on photon subtraction
  using a photon-number-resolving detector at a telecommunications wavelength}.
\newblock
  \href{https://www.doi.org/10.1038/nphoton.2010.158}{\textit{\bibinfo{journal}{Nat.
  Photon.}} \textbf{\bibinfo{volume}{4}}\textbf{,} \bibinfo{pages}{655--660}
  (\bibinfo{year}{2010})}.

\bibitem{gerrits2010}
\bibinfo{author}{Gerrits, T.} \emph{et~al.}
\newblock \bibinfo{title}{Generation of optical coherent-state superpositions
  by number-resolved photon subtraction from the squeezed vacuum}.
\newblock
  \href{https://www.doi.org/10.1103/PhysRevA.82.031802}{\textit{\bibinfo{journal}{Phys.
  Rev. A}} \textbf{\bibinfo{volume}{82}}\textbf{,} \bibinfo{pages}{031802}
  (\bibinfo{year}{2010})}.

\bibitem{yoshikawa2013}
\bibinfo{author}{Yoshikawa, J.-i.}, \bibinfo{author}{Makino, K.},
  \bibinfo{author}{Kurata, S.}, \bibinfo{author}{van Loock, P.} \&
  \bibinfo{author}{Furusawa, A.}
\newblock \bibinfo{title}{Creation, Storage, and On-Demand Release of Optical
  Quantum States with a Negative Wigner Function}.
\newblock
  \href{https://www.doi.org/10.1103/PhysRevX.3.041028}{\textit{\bibinfo{journal}{Phys.
  Rev. X}} \textbf{\bibinfo{volume}{3}}\textbf{,} \bibinfo{pages}{041028}
  (\bibinfo{year}{2013})}.

\bibitem{wang2005}
\bibinfo{author}{Wang, B.} \& \bibinfo{author}{Duan, L.-M.}
\newblock \bibinfo{title}{Engineering superpositions of coherent states in
  coherent optical pulses through cavity-assisted interaction}.
\newblock
  \href{https://www.doi.org/10.1103/PhysRevA.72.022320}{\textit{\bibinfo{journal}{Phys.
  Rev. A}} \textbf{\bibinfo{volume}{72}}\textbf{,} \bibinfo{pages}{022320}
  (\bibinfo{year}{2005})}.

\bibitem{ralph2003}
\bibinfo{author}{Ralph, T.~C.}, \bibinfo{author}{Gilchrist, A.},
  \bibinfo{author}{Milburn, G.~J.}, \bibinfo{author}{Munro, W.~J.} \&
  \bibinfo{author}{Glancy, S.}
\newblock \bibinfo{title}{Quantum computation with optical coherent states}.
\newblock
  \href{https://www.doi.org/10.1103/PhysRevA.68.042319}{\textit{\bibinfo{journal}{Phys.
  Rev. A}} \textbf{\bibinfo{volume}{68}}\textbf{,} \bibinfo{pages}{042319}
  (\bibinfo{year}{2003})}.

\bibitem{gilchrist2004}
\bibinfo{author}{Gilchrist, A.} \emph{et~al.}
\newblock \bibinfo{title}{Schr{\"o}dinger cats and their power for quantum
  information processing}.
\newblock
  \href{https://www.doi.org/10.1088/1464-4266/6/8/032}{\textit{\bibinfo{journal}{J. Opt. B}} \textbf{\bibinfo{volume}{6}}\textbf{,}
  \bibinfo{pages}{S828--S833} (\bibinfo{year}{2004})}.

\bibitem{cochrane1999}
\bibinfo{author}{Cochrane, P.~T.}, \bibinfo{author}{Milburn, G.~J.} \&
  \bibinfo{author}{Munro, W.~J.}
\newblock \bibinfo{title}{Macroscopically distinct quantum-superposition states
  as a bosonic code for amplitude damping}.
\newblock
  \href{https://www.doi.org/10.1103/PhysRevA.59.2631}{\textit{\bibinfo{journal}{Phys.
  Rev. A}} \textbf{\bibinfo{volume}{59}}\textbf{,} \bibinfo{pages}{2631--2634}
  (\bibinfo{year}{1999})}.

\bibitem{leghtas2013}
\bibinfo{author}{Leghtas, Z.} \emph{et~al.}
\newblock \bibinfo{title}{Hardware-Efficient Autonomous Quantum Memory
  Protection}.
\newblock
  \href{https://www.doi.org/10.1103/PhysRevLett.111.120501}{\textit{\bibinfo{journal}{Phys.
  Rev. Lett.}} \textbf{\bibinfo{volume}{111}}\textbf{,} \bibinfo{pages}{120501}
  (\bibinfo{year}{2013})}.

\bibitem{bergmann2016}
\bibinfo{author}{Bergmann, M.} \& \bibinfo{author}{van Loock, P.}
\newblock \bibinfo{title}{Quantum error correction against photon loss using
  multicomponent cat states}.
\newblock
  \href{https://www.doi.org/10.1103/PhysRevA.94.042332}{\textit{\bibinfo{journal}{Phys.
  Rev. A}} \textbf{\bibinfo{volume}{94}}\textbf{,} \bibinfo{pages}{042332}
  (\bibinfo{year}{2016})}.

\bibitem{lund2008}
\bibinfo{author}{Lund, A.~P.}, \bibinfo{author}{Ralph, T.~C.} \&
  \bibinfo{author}{Haselgrove, H.~L.}
\newblock \bibinfo{title}{Fault-Tolerant Linear Optical Quantum Computing with
  Small-Amplitude Coherent States}.
\newblock
  \href{https://www.doi.org/10.1103/PhysRevLett.100.030503}{\textit{\bibinfo{journal}{Phys.
  Rev. Lett.}} \textbf{\bibinfo{volume}{100}}\textbf{,} \bibinfo{pages}{030503}
  (\bibinfo{year}{2008})}.

\bibitem{duan2004}
\bibinfo{author}{Duan, L.-M.} \& \bibinfo{author}{Kimble, H.~J.}
\newblock \bibinfo{title}{Scalable Photonic Quantum Computation through
  Cavity-Assisted Interactions}.
\newblock
  \href{https://www.doi.org/10.1103/PhysRevLett.92.127902}{\textit{\bibinfo{journal}{Phys.
  Rev. Lett.}} \textbf{\bibinfo{volume}{92}}\textbf{,} \bibinfo{pages}{127902}
  (\bibinfo{year}{2004})}.

\bibitem{reiserer2015}
\bibinfo{author}{Reiserer, A.} \& \bibinfo{author}{Rempe, G.}
\newblock \bibinfo{title}{Cavity-based quantum networks with single atoms and
  optical photons}.
\newblock
  \href{https://www.doi.org/10.1103/RevModPhys.87.1379}{\textit{\bibinfo{journal}{Rev.
  Mod. Phys.}} \textbf{\bibinfo{volume}{87}}\textbf{,}
  \bibinfo{pages}{1379--1418} (\bibinfo{year}{2015})}.

\bibitem{jeong2002}
\bibinfo{author}{Jeong, H.} \& \bibinfo{author}{Kim, M.~S.}
\newblock \bibinfo{title}{Efficient quantum computation using coherent states}.
\newblock
  \href{https://www.doi.org/10.1103/PhysRevA.65.042305}{\textit{\bibinfo{journal}{Phys.
  Rev. A}} \textbf{\bibinfo{volume}{65}}\textbf{,} \bibinfo{pages}{042305}
  (\bibinfo{year}{2002})}.

\bibitem{schleich1991}
\bibinfo{author}{Schleich, W.}, \bibinfo{author}{Pernigo, M.} \&
  \bibinfo{author}{Kien, F.~L.}
\newblock \bibinfo{title}{Nonclassical state from two pseudoclassical states}.
\newblock
  \href{https://www.doi.org/10.1103/PhysRevA.44.2172}{\textit{\bibinfo{journal}{Phys.
  Rev. A}} \textbf{\bibinfo{volume}{44}}\textbf{,} \bibinfo{pages}{2172--2187}
  (\bibinfo{year}{1991})}.

\bibitem{dariano1995}
\bibinfo{author}{D'Ariano, G.~M.}, \bibinfo{author}{Leonhardt, U.} \&
  \bibinfo{author}{Paul, H.}
\newblock \bibinfo{title}{Homodyne detection of the density matrix of the
  radiation field}.
\newblock
  \href{https://www.doi.org/10.1103/PhysRevA.52.R1801}{\textit{\bibinfo{journal}{Phys.
  Rev. A}} \textbf{\bibinfo{volume}{52}}\textbf{,}
  \bibinfo{pages}{R1801--R1804} (\bibinfo{year}{1995})}.

\bibitem{buzek1992}
\bibinfo{author}{Bu\ifmmode~\check{z}\else \v{z}\fi{}ek, V.},
  \bibinfo{author}{Vidiella-Barranco, A.} \& \bibinfo{author}{Knight, P.~L.}
\newblock \bibinfo{title}{Superpositions of coherent states: Squeezing and
  dissipation}.
\newblock
  \href{https://www.doi.org/10.1103/PhysRevA.45.6570}{\textit{\bibinfo{journal}{Phys.
  Rev. A}} \textbf{\bibinfo{volume}{45}}\textbf{,} \bibinfo{pages}{6570--6585}
  (\bibinfo{year}{1992})}.

\bibitem{spagnolo2009}
\bibinfo{author}{Spagnolo, N.}, \bibinfo{author}{Vitelli, C.},
  \bibinfo{author}{De~Angelis, T.}, \bibinfo{author}{Sciarrino, F.} \&
  \bibinfo{author}{De~Martini, F.}
\newblock \bibinfo{title}{Wigner-function theory and decoherence of the
  quantum-injected optical parametric amplifier}.
\newblock
  \href{https://www.doi.org/10.1103/PhysRevA.80.032318}{\textit{\bibinfo{journal}{Phys.
  Rev. A}} \textbf{\bibinfo{volume}{80}}\textbf{,} \bibinfo{pages}{032318}
  (\bibinfo{year}{2009})}.

\bibitem{vlastakis2015}
\bibinfo{author}{Vlastakis, B.} \emph{et~al.}
\newblock \bibinfo{title}{Characterizing entanglement of an artificial atom and
  a cavity cat state with Bell's inequality}.
\newblock
  \href{https://www.doi.org/10.1038/ncomms9970}{\textit{\bibinfo{journal}{Nat.
  Commun.}} \textbf{\bibinfo{volume}{6}}\textbf{,} \bibinfo{pages}{8970}
  (\bibinfo{year}{2015})}.

\bibitem{vidal2002}
\bibinfo{author}{Vidal, G.} \& \bibinfo{author}{Werner, R.~F.}
\newblock \bibinfo{title}{Computable measure of entanglement}.
\newblock
  \href{https://www.doi.org/10.1103/PhysRevA.65.032314}{\textit{\bibinfo{journal}{Phys.
  Rev. A}} \textbf{\bibinfo{volume}{65}}\textbf{,} \bibinfo{pages}{032314}
  (\bibinfo{year}{2002})}.

\bibitem{nielsen2000}
\bibinfo{author}{Nielsen, M.~A.} \& \bibinfo{author}{Chuang, I.~L.}
\newblock \emph{\bibinfo{title}{Quantum {Computation} and {Quantum}
  {Information}}} (\bibinfo{publisher}{Cambridge University Press},
  \bibinfo{year}{2000}).

\bibitem{ofek2016}
\bibinfo{author}{Ofek, N.} \emph{et~al.}
\newblock \bibinfo{title}{Extending the lifetime of a quantum bit with error
  correction in superconducting circuits}.
\newblock
  \href{https://www.doi.org/10.1038/nature18949}{\textit{\bibinfo{journal}{Nature}}
  \textbf{\bibinfo{volume}{536}}\textbf{,} \bibinfo{pages}{441--445}
  (\bibinfo{year}{2016})}.

\bibitem{kimble2008}
\bibinfo{author}{Kimble, H.~J.}
\newblock \bibinfo{title}{The quantum internet}.
\newblock
  \href{https://www.doi.org/10.1038/nature07127}{\textit{\bibinfo{journal}{Nature}}
  \textbf{\bibinfo{volume}{453}}\textbf{,} \bibinfo{pages}{1023--1030}
  (\bibinfo{year}{2008})}.

\bibitem{teo2013}
\bibinfo{author}{Teo, C.} \emph{et~al.}
\newblock \bibinfo{title}{Realistic loophole-free {Bell} test with
  atom{\textendash}photon entanglement}.
\newblock
  \href{https://www.doi.org/10.1038/ncomms3104}{\textit{\bibinfo{journal}{Nat.
  Commun.}} \textbf{\bibinfo{volume}{4}}\textbf{,} \bibinfo{pages}{2104}
  (\bibinfo{year}{2013})}.

\bibitem{kwon2013}
\bibinfo{author}{Kwon, H.} \& \bibinfo{author}{Jeong, H.}
\newblock \bibinfo{title}{Violation of the Bell--Clauser-Horne-Shimony-Holt
  inequality using imperfect photodetectors with optical hybrid states}.
\newblock
  \href{https://www.doi.org/10.1103/PhysRevA.88.052127}{\textit{\bibinfo{journal}{Phys.
  Rev. A}} \textbf{\bibinfo{volume}{88}}\textbf{,} \bibinfo{pages}{052127}
  (\bibinfo{year}{2013})}.

\bibitem{kalb2015}
\bibinfo{author}{Kalb, N.}, \bibinfo{author}{Reiserer, A.},
  \bibinfo{author}{Ritter, S.} \& \bibinfo{author}{Rempe, G.}
\newblock \bibinfo{title}{Heralded Storage of a Photonic Quantum Bit in a
  Single Atom}.
\newblock
  \href{https://www.doi.org/10.1103/PhysRevLett.114.220501}{\textit{\bibinfo{journal}{Phys.
  Rev. Lett.}} \textbf{\bibinfo{volume}{114}}\textbf{,} \bibinfo{pages}{220501}
  (\bibinfo{year}{2015})}.

\bibitem{andersen2015}
\bibinfo{author}{Andersen, U.~L.}, \bibinfo{author}{Neergaard-Nielsen, J.~S.},
  \bibinfo{author}{Van~Loock, P.} \& \bibinfo{author}{Furusawa, A.}
\newblock \bibinfo{title}{Hybrid discrete-and continuous-variable quantum
  information}.
\newblock
  \href{https://www.doi.org/10.1038/nphys3410}{\textit{\bibinfo{journal}{Nat.
  Phys.}} \textbf{\bibinfo{volume}{11}}\textbf{,} \bibinfo{pages}{713}
  (\bibinfo{year}{2015})}.

\end{thebibliography}

\clearpage





\begin{thebibliography}{S9}

\bibitem[44]{thompson1992}
\bibinfo{author}{Thompson, R.~J.}, \bibinfo{author}{Rempe, G.} \&
  \bibinfo{author}{Kimble, H.~J.}
\newblock \bibinfo{title}{Observation of Normal-Mode Splitting for an Atom in
  an Optical Cavity}.
\newblock
  \href{https://www.doi.org/10.1103/PhysRevLett.68.1132}{\textit{\bibinfo{journal}{Phys.
  Rev. Lett.}} \textbf{\bibinfo{volume}{68}}\textbf{,}
  \bibinfo{pages}{1132--1135} (\bibinfo{year}{1992})}.
  
\bibitem[45]{lvovsky2004}
\bibinfo{author}{Lvovsky, A.~I.}
\newblock \bibinfo{title}{Iterative maximum-likelihood reconstruction in
  quantum homodyne tomography}.
\newblock
  \href{https://www.doi.org/10.1088/1464-4266/6/6/014}{\textit{\bibinfo{journal}{J
  Opt B Quantum Semiclassical Opt}} \textbf{\bibinfo{volume}{6}}\textbf{,}
  \bibinfo{pages}{S556--559} (\bibinfo{year}{2004})}.

\bibitem[46]{banaszek1999}
\bibinfo{author}{Banaszek, K.}, \bibinfo{author}{D'Ariano, G.~M.},
  \bibinfo{author}{Paris, M. G.~A.} \& \bibinfo{author}{Sacchi, M.~F.}
\newblock \bibinfo{title}{Maximum-likelihood estimation of the density matrix}.
\newblock
  \href{https://www.doi.org/10.1103/PhysRevA.61.010304}{\textit{\bibinfo{journal}{Phys.
  Rev. A}} \textbf{\bibinfo{volume}{61}}\textbf{,}
  \bibinfo{pages}{010304} (\bibinfo{year}{1999})}.

\bibitem[47]{kuhn2015}
\bibinfo{author}{Kuhn, A.}
\newblock \href{https://www.doi.org/10.1007/978-3-319-19231-4_1}{
  \bibinfo{title}{Cavity Induced Interfacing of Atoms and Light}}.
\newblock In \textit{\bibinfo{booktitle}{Engineering the Atom--Photon Interaction}}
  (eds \bibinfo{editor}{Predojevi{\'c}, A.} \&
  \bibinfo{editor}{Mitchell, M.~W.})
  \bibinfo{pages}{3--38} (\bibinfo{publisher}{Springer, Cham},
  \bibinfo{year}{2015}).

\end{thebibliography}

\clearpage


\begin{thebibliography}{S9}

\bibitem[S1]{appel2007_supp}
\bibinfo{author}{Appel, J.}, \bibinfo{author}{Hoffman, D.},
  \bibinfo{author}{Figueroa, E.} \& \bibinfo{author}{Lvovsky, A.~I.}
\newblock \emph{\bibinfo{title}{Electronic noise in optical homodyne
  tomography}}.
\newblock
  \href{https://www.doi.org/10.1103/PhysRevA.75.035802}{\textit{\bibinfo{journal}{Phys.
  Rev. A}} \textbf{\bibinfo{volume}{75}}\textbf{,} \bibinfo{pages}{035802}
  (\bibinfo{year}{2007})}.

\bibitem[S2]{yurke1986_supp}
\bibinfo{author}{Yurke, B.} \& \bibinfo{author}{Stoler, D.}
\newblock \emph{\bibinfo{title}{Generating quantum mechanical superpositions of
  macroscopically distinguishable states via amplitude dispersion}}.
\newblock
  \href{https://www.doi.org/10.1103/PhysRevLett.57.13}{\textit{\bibinfo{journal}{Phys.
  Rev. Lett.}} \textbf{\bibinfo{volume}{57}}\textbf{,} \bibinfo{pages}{13--16}
  (\bibinfo{year}{1986})}.

\bibitem[S3]{duan2004_supp}
\bibinfo{author}{Duan, L.-M.} \& \bibinfo{author}{Kimble, H.~J.}
\newblock \emph{\bibinfo{title}{Scalable Photonic Quantum Computation through
  Cavity-Assisted Interactions}}.
\newblock
  \href{https://www.doi.org/10.1103/PhysRevLett.92.127902}{\textit{\bibinfo{journal}{Phys.
  Rev. Lett.}} \textbf{\bibinfo{volume}{92}}\textbf{,} \bibinfo{pages}{127902}
  (\bibinfo{year}{2004})}.

\bibitem[S4]{kalb2015_supp}
\bibinfo{author}{Kalb, N.}, \bibinfo{author}{Reiserer, A.},
  \bibinfo{author}{Ritter, S.} \& \bibinfo{author}{Rempe, G.}
\newblock \emph{\bibinfo{title}{Heralded Storage of a Photonic Quantum Bit in a
  Single Atom}}.
\newblock
  \href{https://www.doi.org/10.1103/PhysRevLett.114.220501}{\textit{\bibinfo{journal}{Phys.
  Rev. Lett.}} \textbf{\bibinfo{volume}{114}}\textbf{,} \bibinfo{pages}{220501}
  (\bibinfo{year}{2015})}.

\bibitem[S5]{heeres2017_supp}
\bibinfo{author}{Heeres, R.~W.} \emph{et~al.}
\newblock \emph{\bibinfo{title}{Implementing a universal gate set on a logical
  qubit encoded in an oscillator}}.
\newblock
  \href{https://www.doi.org/10.1038/s41467-017-00045-1}{\textit{\bibinfo{journal}{Nat.
  Commun.}} \textbf{\bibinfo{volume}{8}}\textbf{,} \bibinfo{pages}{94}
  (\bibinfo{year}{2017})}.

\bibitem[S6]{leghtas2013_supp}
\bibinfo{author}{Leghtas, Z.} \emph{et~al.}
\newblock \emph{\bibinfo{title}{Hardware-Efficient Autonomous Quantum Memory
  Protection}}.
\newblock
  \href{https://www.doi.org/10.1103/PhysRevLett.111.120501}{\textit{\bibinfo{journal}{Phys.
  Rev. Lett.}} \textbf{\bibinfo{volume}{111}}\textbf{,} \bibinfo{pages}{120501}
  (\bibinfo{year}{2013})}.

\end{thebibliography}
\end{document}